\documentclass[2column]{aa}
\usepackage{natbib}
\usepackage{multirow}
\usepackage{color}
\usepackage[colorlinks=true]{hyperref}
\hypersetup{citecolor= blue, linkcolor=red, urlcolor=blue}

\usepackage[us,12hr]{datetime}
\usepackage{ulem}
\definecolor{deepmagenta}{rgb}{0.8, 0.0, 0.8}
\definecolor{lightblue}{rgb}{0.2, 0.4, 0.8}
\definecolor{forestgreen}{RGB}{34, 139, 34}

\begin{document}

   \title{Stratification of physical parameters in a C-class solar flare using multiline observations }
   \author{Rahul Yadav
          \inst{1}
          \and
          C. J. D\'iaz Baso
          \inst{1}
          \and
         J. de la Cruz Rodr\'iguez
          \inst{1}
          \and 
          Flavio Calvo
          \inst{1}
            \and 
          Roberta Morosin
          \inst{1}}

   \institute{Institute for Solar Physics, Dept. of Astronomy, Stockholm University, AlbaNova University Centre, SE-10691 Stockholm, Sweden
              \email{rahul.yadav@astro.su.se}
             }

   \date{Draft: compiled on \today\ at \currenttime~UT}
   
   \authorrunning{Yadav et al.}
   \titlerunning{}
 \abstract{
We present high-resolution and multiline observations of a C2-class solar flare (SOL2019-05-06T08:47), which occurred in NOAA AR 12740 on May 6, 2019. 
The rise, peak, and decay phases of the flare were recorded continuously and quasi-simultaneously in the \ion{Ca}{ii}~K line with the CHROMIS instrument and in the \ion{Ca}{ii}~8542~\AA~and \ion{Fe}{i}~6173~\AA~lines with the CRISP instrument at the Swedish 1-m Solar Telescope. The observations in the chromospheric \ion{Ca}{ii}~lines exhibit intense brightening near the flare footpoints. At these locations, a nonlocal thermodynamic equilibrium inversion code was employed to infer the temperature, magnetic field, line-of-sight (LOS) velocity, and microturbulent velocity stratification in the flaring atmosphere.  The temporal analysis of the inferred temperature at the flare footpoints shows that the flaring atmosphere from $\log\tau_{500}\sim-2.5$ to $-3.5$ is heated up to 7~kK, whereas from $\log\tau_{500}\sim-3.5$ to $-5$ the inferred temperature ranges between $\sim7.5$~kK and $\sim11$~kK.  During the flare peak time, the LOS velocity shows both upflows and downflows around the flare footpoints in the upper chromosphere and lower chromosphere, respectively. Moreover, the temporal analysis of the LOS magnetic field at the flare points exhibits a maximum change of $\sim$ 600~G. After the flare, the LOS magnetic field decreases to the non-flaring value, exhibiting no permanent or step-wise change. The analysis of response functions to the temperature, LOS magnetic field, and velocity shows that the \ion{Ca}{ii} lines exhibit enhanced sensitivity to the deeper layers (i.e., $\log\tau_{500}\sim-3$) of the flaring atmosphere, whereas for the non-flaring atmosphere they are mainly sensitive around $\log\tau_{500}\sim-4$. We suggest that a fraction of the apparent increase in the LOS magnetic field at the flare footpoints may be due to the increase in the sensitivity of the \ion{Ca}{ii}~8542~\AA\ line in the deeper layers, where the field strength is relatively strong. The rest may be due to magnetic field reconfiguration during the flare. In the photosphere, we do not notice significant changes in the physical parameters during the flare or non-flare times. Our observations illustrate that even a less intense C-class flare can heat the deeper layers of the solar chromosphere, mainly at the flare footpoints, without affecting the photosphere.
}

\maketitle

\section{Introduction}
Solar flares are sudden energy release phenomena that last from a few minutes to hours, resulting in plasma heating, particle acceleration, and energy release in the entire electromagnetic spectrum. A significant amount of energy, up to 10$^{32}$ erg, is released in an intense solar flare. It is believed that they are caused by the magnetic reconnection or re-configuration of magnetic field lines in the corona, converting the magnetic energy into the kinematic and thermal energy.
For an overview of solar flares, we refer to \cite{2011SSRv..159...19F}, \cite{ 2011LRSP....8....6S}, \cite{2015SoPh..290.3425J}, and \cite{2017LRSP...14....2B}.

It is known that a major amount of energy is transported to the dense chromosphere along the coronal magnetic loops by accelerated particles or thermal conduction or in the form of waves, although the processes that describe this transfer are still a matter of debate (e.g., \citealt{2016ApJ...827..101K}). Several decades of observational and theoretical investigations performed by various authors have revealed a number of processes and mechanisms. 
For example, in the collisional thick-target model (CTTM), the energetic particles produced at the reconnection site in the corona deposit their energy via Coulomb collisions in the chromosphere. This deposition of energy gives rise to intense brightening and emission of hard and soft X-rays near the location of the coronal magnetic footpoints \citep{1971SoPh...18..489B,1972SoPh...24..414H}. 
In addition to this, the transfer of energy to the lower atmosphere via Alfv\'en waves was first proposed by \cite{1982SoPh...80...99E}. Later, \cite{2008ApJ...675.1645F}  demonstrated that Alfv\'enic perturbations in the coronal field lines can supply an adequate amount of energy to heat the chromosphere through various wave dissipation mechanisms. Recently, using radiation hydrodynamic simulations, \cite{2016ApJ...827..101K} demonstrated that both mechanisms (Alfv\'en wave dissipation and electron beam collisional losses) are capable of producing strong chromospheric heating.

Several observations have shown that a large amount of the flare energy is radiated in the chromosphere. Therefore, this layer is of great importance and is key for understanding how and where the flare energy is transported and dissipated. Some of the most common chromospheric lines utilized to investigate flares  are H$\alpha$, \ion{Ca}{ii}~H \& K, \ion{Ca}{ii}~8542~\AA,~and \ion{He}{i}~10830~\AA. These spectral lines have been used to construct semiempirical models of a flaring atmosphere by several authors \citep{1975SoPh...42..395M,2002A&A...387..678F,2014A&A...561A..98S,Judge2014,2017ApJ...834...26K,2017ApJ...846....9K,Kuridze2018,2019A&A...621A..35L,2020arXiv200901537V}.

In addition to other physical parameters, the magnetic field plays a crucial role in the triggering and development of a flare \citep{2015ApJ...798..135B}. Thus, for a comprehensive modeling of flares, spectropolarimetric observations of spectral lines that sample different layers of the solar atmosphere are needed to infer the stratification of the magnetic field. 
The magnetic field changes in the photosphere during strong flares have been investigated by several authors as they have been commonly observed from various space- and ground-based observations \citep{2005ApJ...635..647S,2010ApJ...724.1218P,2012ApJ...747..134M,2018ApJ...852...25C}. In a recent statistical analysis of 75 solar flares, \cite{2018ApJ...852...25C} found that most of the intense flares (>M1.6-class) exhibit abrupt and permanent magnetic field changes in the photosphere.  
To investigate the presence of such changes in the chromosphere, observations with exceedingly high polarimetric  sensitivity are required. 
However, chromospheric polarimetry during a flare is challenging due to a generally lower magnetic field strength in the chromosphere, the scarcity of sensitive spectral lines, the seeing-induced limitations, instrumental constraints, and the flare's unpredictability. As a result, we have limited flare studies that were performed with the chromospheric polarimetry. Therefore, it is still not fully understood how the magnetic field or other physical parameters behave in the chromosphere during the different stages of a flare.

The \ion{Ca}{ii}~8542~\AA\ line has been one of the most studied lines in investigations of the magnetic, thermodynamic, and kinematic structure of the chromosphere (e.g., \citealt{2007ApJ...670..885P,2008A&A...480..515C,2012A&A...543A..34D, 2012ApJ...748..138K, 2016MNRAS.459.3363Q, 2020A&A...637A...1K, 2020arXiv200614486P}). Spectropolarimetric observations of flares in the \ion{Ca}{ii}~8542~\AA~line are now available from different ground-based instruments, for example the Interferometric Bidimensional Spectropolarimeter \citep[IBIS;] []{2006SoPh..236..415C} and the CRisp Imaging SpectroPolarimeter \citep[CRISP;] []{Scharmer2008}. The available inversion methods allow us to infer the physical parameters of the solar atmosphere from such polarimetric observations (see \citealt{2017SSRv..210..109D} and references therein). 

Recently, \cite{2017ApJ...834...26K} detected a first direct step-wise change during a X1-class flare in the chromospheric line-of-sight (LOS) magnetic field through spectropolarimetry of the \ion{Ca}{ii}~8542~\AA\ line using the weak-field approximation (WFA; \citealt{2004ASSL..307.....L}). They reported stronger field changes in the chromosphere compared to their photospheric counterparts.
Using a nonlocal thermodynamic equilibrium (non-LTE) inversion of the \ion{Ca}{ii}~8542~\AA~line, \cite{2017ApJ...846....9K} found that during a C8.4-class flare, the temperature in the middle and upper chromosphere is enhanced from $\sim6.5$~kK to $\sim20$~kK between log~$\tau \sim~-3.5$ and $-5.5$ in the flaring atmosphere, compared to a pre-flare temperature of $\sim5 - 10$~kK. They also reported that the flaring chromosphere is dominated by downflowing condensations at the formation height of the \ion{Ca}{ii}~8542~\AA~line. Furthermore, \cite{Kuridze2018} constructed a semiempirical model of the flaring atmosphere from the spectropolarimetric inversion of the \ion{Ca}{ii}~8542~\AA~line using the non-LTE NICOLE code \citep{2015A&A...577A...7S}. They reported that during the flare peak time the polarization signals become stronger when compared to the pre-flare and post-flare profiles. They also showed that the \ion{Ca}{ii}~8542~\AA~line is more sensitive to the lower atmosphere during the flaring time.

Generally, a flare influences different layers of the solar atmosphere almost simultaneously. Therefore, a single spectral line may not encode sufficient information of all the relevant physical parameters to fully characterize a flare. A more reliable flaring atmosphere can be constructed from the simultaneous investigation of multiple spectral lines forming in different parts of the solar atmosphere. In this paper, we present a multiline investigation of a C2-class flare, which is observed simultaneously in three spectral lines (\ion{Ca}{ii}~K, \ion{Ca}{ii}~8542~\AA,~and \ion{Fe}{i}~6173~\AA). More importantly, we also have simultaneous spectropolarimetric observations in the photosphere (\ion{Fe}{i}~6173~\AA) and the chromosphere (\ion{Ca}{ii}~8542~\AA). Such spectropolarimetric observations mapping two different layers can be utilized to understand the magnetic field topology and coupling of the solar atmosphere (e.g., \citealt{2019A&A...632A.112Y}).

We employed a non-LTE inversion code to infer the atmospheric stratification of the temperature, LOS velocity, microturbulent velocity, and magnetic field in the flaring atmosphere, making use of all three lines simultaneously.
The purpose of this study is to investigate the temporal behavior of the inferred parameters, mainly at the flare footpoints, in the solar atmosphere during flare and non-flare times.

This article is organized in the following manner. The overview of observations and flare are given in Sections \ref{sec-observation} and \ref{overview}, respectively. The analysis of spectropolarimetric data is discussed in Sect.~\ref{sec-inversion}. The obtained results are described in Sect.~\ref{sec_results}. Finally, the paper is summarized in Sect.~\ref{sec_conclusion}. 

\begin{figure}[!t]
\centering
\includegraphics[width=0.4\textwidth]{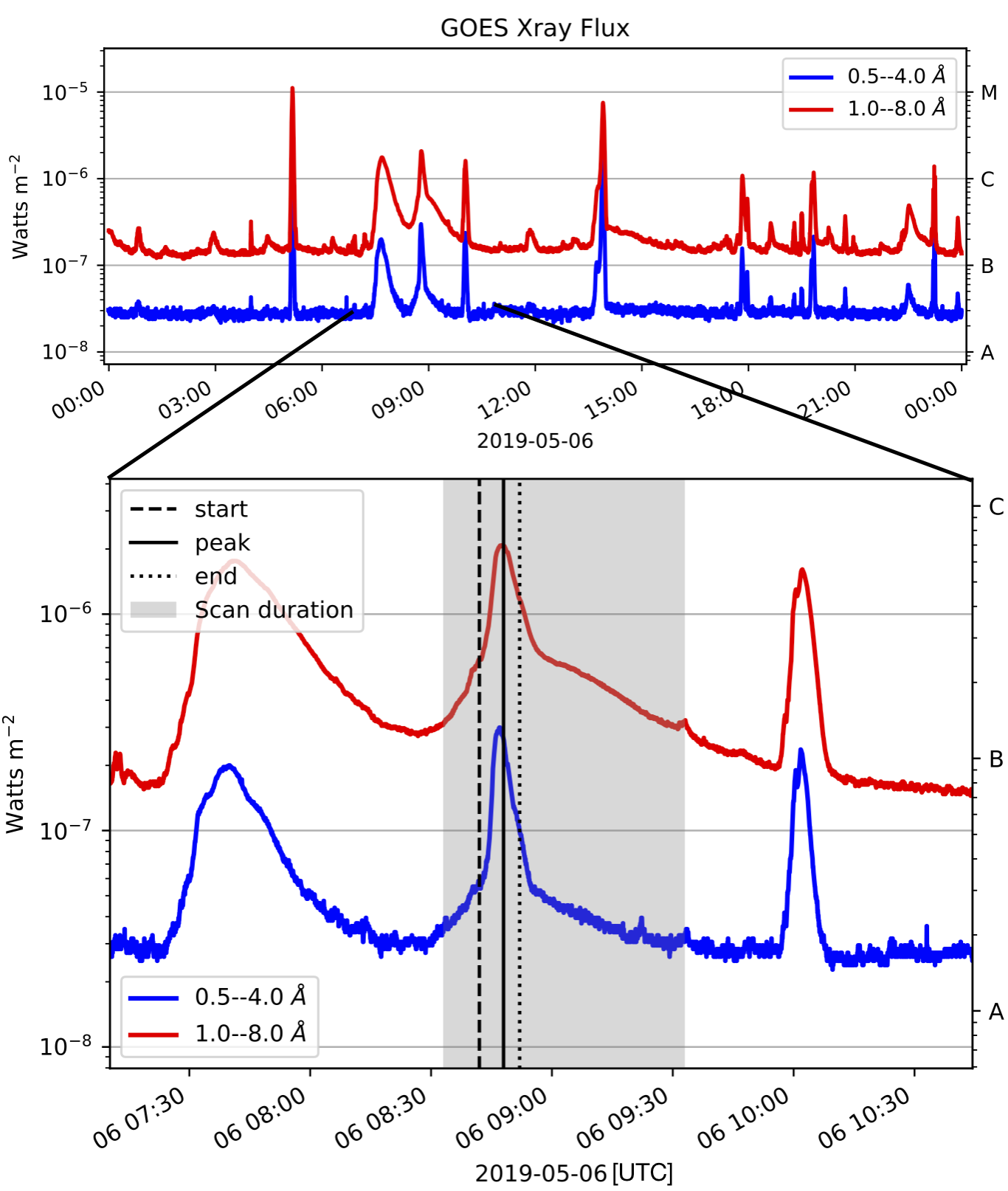}
\caption{\footnotesize{Time evolution of GOES X-ray flux observed on May 6, 2019. The vertical dashed, solid, and dotted lines refer to the start, peak, and end times of the flare, respectively. The gray shaded area indicates the observing window with the SST.}}
\label{fig_goes_flux}

\end{figure}

\begin{figure*}[!t]
\centering
\includegraphics[width=1\textwidth]{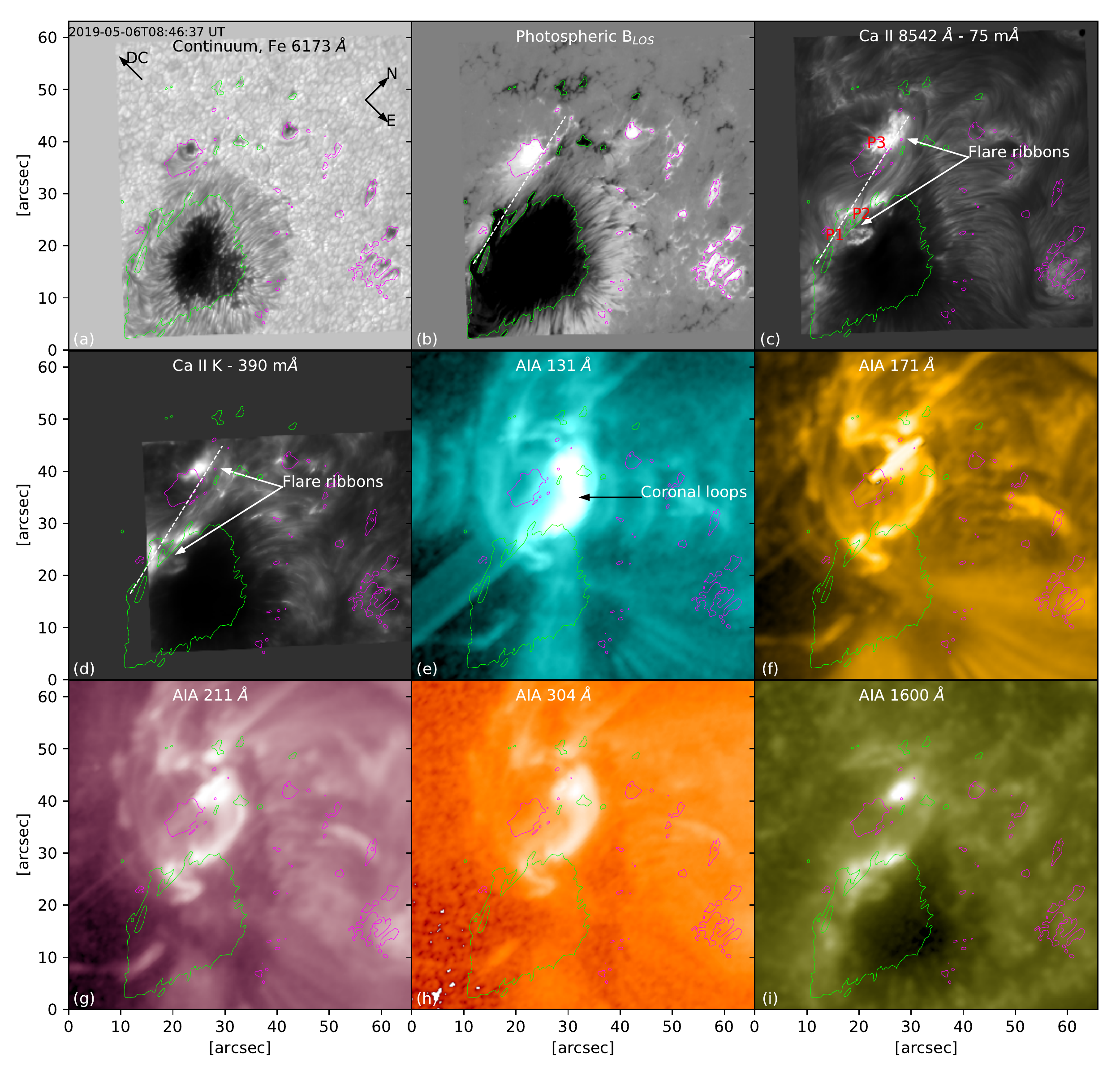}

\caption{\footnotesize Overview of the C2-class flare observed at 08:45:55~UT on May 6, 2019. (a) Continuum map showing the sunspot and the pores. (b) Line-of-sight magnetic field as inferred from a Milne-Eddington inversion of the \ion{Fe}{i} 6173.34~\AA~line. The color scale extends in the $\pm{0.8}$~kG range, where black and white represent the negative and positive polarity, respectively. (c - d) Chromospheric intensity maps, observed with the CRISP and CHROMIS instruments at the SST, showing bright flare ribbons at the flare footpoints. (e -- i) AIA images observed in different channels, in logarithmic scale. The green and magenta contours represent the negative and positive polarity in the photosphere, respectively. A dashed white line and the ``P1,'' ``P2,'' and ``P3'' in panel~(c) highlight the location of the pixels analyzed in the paper. Solar north, solar east, and the direction of the disk center are indicated by ``N,'' ``E,'' and ``DC,'' respectively.}
\label{fig_overview}
\end{figure*}

\section{Observations and data reduction}
\label{sec-observation}
Observations of the active region (AR) NOAA~12740, located at 
N08E48 ($\mu$=0.62), were recorded between 08:34 and 9:33 UT on May 6, 2019, with the CRisp Imaging SpectroPolarimeter \citep[CRISP;] []{Scharmer2008} and the CHROMospheric Imaging Spectrometer \citep[CHROMIS;][]{2017psio.confE..85S} instruments at the Swedish 1-m Solar Telescope \citep[SST;][]{2003SPIE.4853..341S}. Both instruments are equipped with a dual Fabry-P\'erot tunable filter, but CRISP has additional polarimetric capabilities. The CRISP simultaneously recorded full spectropolarimetric data in the \ion{Ca}{ii} 8542 \AA~and \ion{Fe}{i}~6173 \AA~spectral lines. The \ion{Ca}{ii} 8542 \AA~line scans consisted of 17 wavelength positions spanning a range of 1.4~\AA\ around line center, with steps of 75~m\AA~in the inner wings and 125~m\AA~at four outer wing positions, whereas the \ion{Fe}{i}~6173~\AA\ spectral line was scanned at 15 wavelength positions spanning a range of 0.5~\AA\ around line center, with steps of 35 m\AA~in the inner wings and 50 m\AA~in the outer wings at four different positions. The CRISP data were obtained with a cadence of 21~sec with a pixel scale of 0.057\arcsec. The CHROMIS recorded \ion{Ca}{ii}~K intensity profiles at 28 wavelength positions spanning a range of 3~\AA\ around line center, with steps of 65 m\AA\ in the inner wings and an additional sampling at wavelength positions $\pm{1.5}$~\AA\ and $\pm{1.18}$~\AA\ relative to the line center. In addition to this, one point in the continuum at 4000~\AA\ was also observed with the CHROMIS instrument. The CHROMIS data have a cadence of 15~sec and a pixel size of 0.0375\arcsec.

The data obtained with CRISP were reduced using the CRISPRED \citep{2015A&A...573A..40D} post-processing pipeline, which includes image reconstruction through multi-object multi-frame blind deconvolution (MOMFBD; \citealt{2005SoPh..228..191V}) and removal of small-scale seeing-induced deformations. The CHROMIS data were reduced using the CHROMISRED pipeline \citep{2018arXiv180403030L}. The CRISP data were aligned with the CHROMIS data and resampled to the CHROMIS pixel scale of 0.0375\arcsec. As the CRISP data were obtained with a lower cadence, we interpolated the CRISP data to the CHROMIS cadence using nearest-neighbor interpolation. For all data, the intensity calibration was performed with the quiet Sun data located close to the disk center after taking into account the limb darkening, whereas the absolute wavelengths were calibrated with the atlas profiles given by \cite{1984SoPh...90..205N}.
In order to improve the signal-to-noise of the polarization signals in the \ion{Ca}{ii} 8542 \AA\ line, we applied the denoising neural network described in \cite{2019A&A...629A..99D}. We checked that the noise reduction process does not affect the strength of the signals.
Furthermore, we performed $2\times2$ pixel spatial binning to enhance the signal-to-noise ratio, which resulted in a final resolution of 0.11\arcsec\ per pixel.

In addition to the SST data, we also analyzed ultraviolet (UV) and extreme ultraviolet (EUV) images observed by the Atmosphere Imaging Assembly (AIA; \citealt{2012SoPh..275...17L}), as well as full-disk continuum images and LOS magnetograms from the Helioseismic and Magnetic Imager (HMI; \citealt{2012SoPh..275..207S}) aboard the Solar Dynamic Observatory (SDO; \cite{2012SoPh..275..207S}). The AIA takes full-disk images in seven EUV bands with a cadence of 12~sec and in two UV bands at 1600~\AA\ and 1700~\AA\ with a cadence of 24~sec. The spatial scale of AIA images is 0.6\arcsec per pixel. All AIA and HMI images were corrected using the standard solar software (SSW) routines (e.g., \texttt{aia\_prep.pro} and \texttt{hmi\_prep.pro}). Finally, all AIA, HMI, CRISP, and CHROMIS data were co-aligned using an image cross-correlation approach. We note that the spatial resolution and temporal cadence of the SDO and the SST data are different.

\section{Overview of the flare}
\label{overview}
A solar flare is generally classified using the one-to-eight scale of$\AA$~soft X-ray peak intensities, which are measured by the Geostationary Operational Environmental Satellite (GOES). The GOES X-ray flux has revealed that AR~12740 produced several C- and M-class flares (14~C-class and one~M-class) during its transit through the solar disk. The flare (SOL2019-05-06T08:47) observed with the SST is classified by GOES as a C2 event. As illustrated in Fig. \ref{fig_goes_flux}, this flare started at 08:41~UT, with its peak and stop times at 08:47~UT and 08:51~UT, respectively. Luckily, this event was recorded under stable seeing conditions from 08:34~UT to 09:33~UT at the SST, covering the rise, peak, and decay phases of the flare. Figure \ref{fig_overview} displays the appearance of the flare in different channels of the AIA instrument as well as the SST data near the flare peak time. The photospheric magnetograms derived from the CRISP \ion{Fe}{i} line demonstrate that the flare originated from a complex magnetic field configuration (see Fig. \ref{fig_overview}b). The observed AR consists of a sunspot and a small pore with opposite polarity. The temporal analysis reveals the presence of a small-scale flux-emerging region between the penumbra and the pore region.
This new flux-emerging region may increase the shearing and can create strong magnetic field gradients, which might destabilize the magnetic field and may be responsible for the onset of the present flare (\citealt{2017LRSP...14....2B}, and references therein).

Figures \ref{fig_overview}c--d show the appearance of enhanced chromospheric intensity around flare footpoints in the field-of-view (FOV) of CRISP and CHROMIS. 
One of the footpoints lies close to the penumbral region of the sunspot, whereas another one is located near a pore of opposite polarity. Remarkably, as shown in Fig.~\ref{fig_overview}, the counterpart of the flare footpoints are also present in the 1600~\AA~channel of the AIA as intense brightenings. In contrast to the chromospheric observations, we do not notice any significant brightening in the photospheric observations.

Furthermore, we observe that the atmosphere above the photosphere is dynamic, and intense brightening is apparent  near flare ribbons in all AIA/SDO channels. Images in EUV 131~\AA, 171~\AA, 211~\AA ,~and 304~\AA~display the hot coronal flaring loop connecting the two footpoints of opposite polarity. The intensity saturation in 131~\AA~indicates the presence of hot plasma ($\sim$1~Mk) in the flaring loop (see Fig.~\ref{fig_overview}e).

\begin{figure*}[!t]
\centering
\includegraphics[width=.325\linewidth]{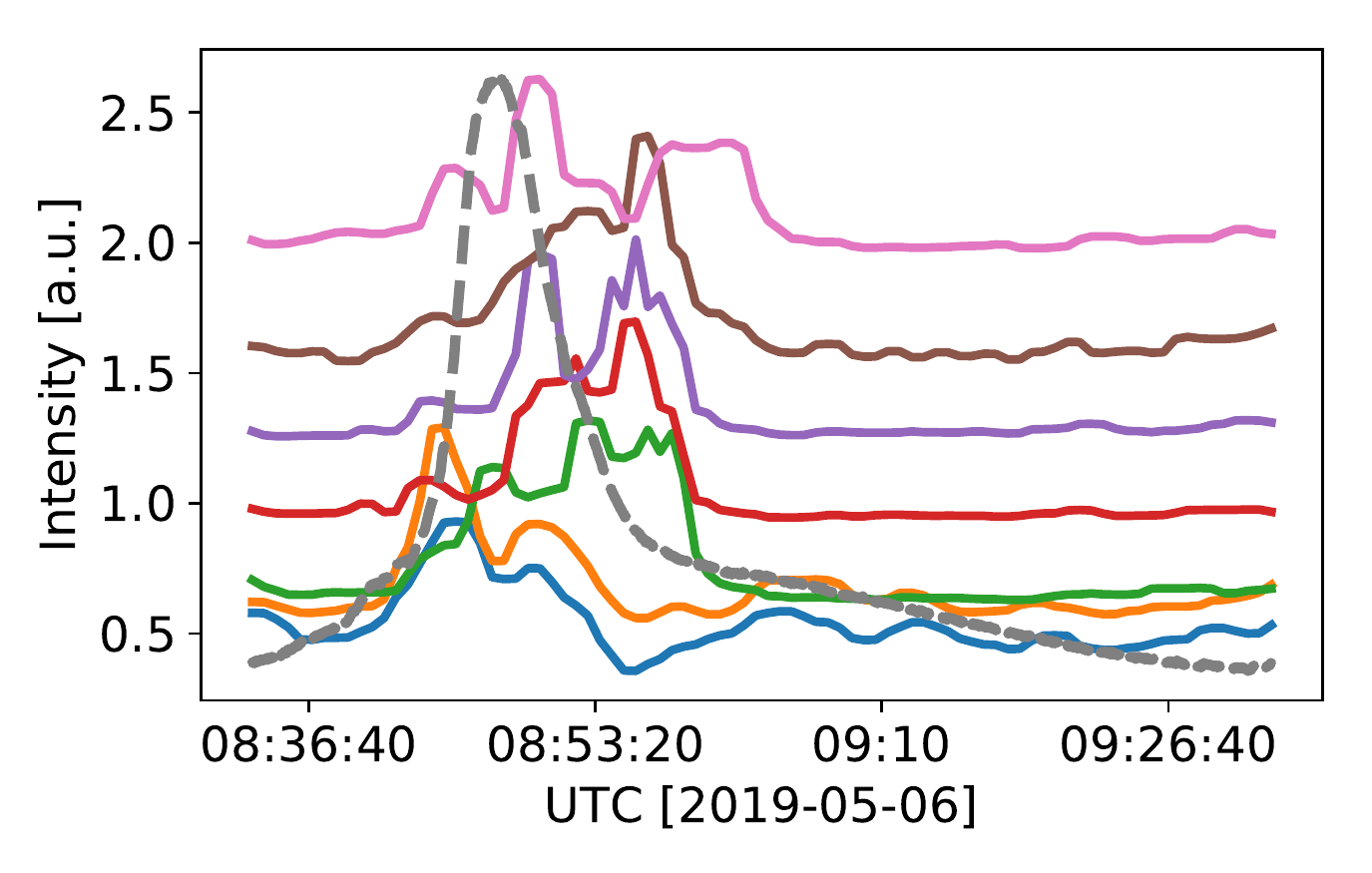}
\includegraphics[width=.325\linewidth]{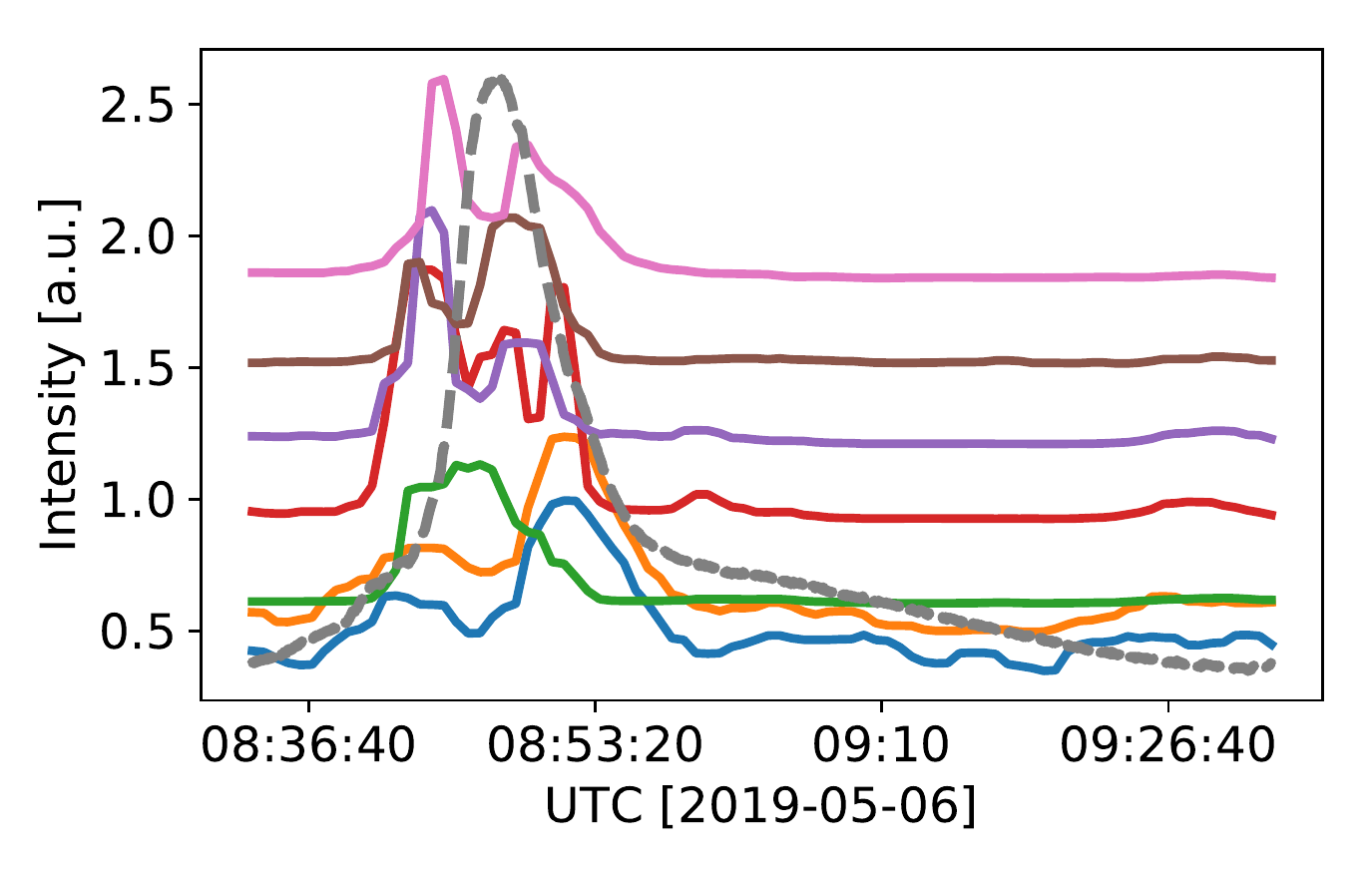}
\includegraphics[width=.325\linewidth]{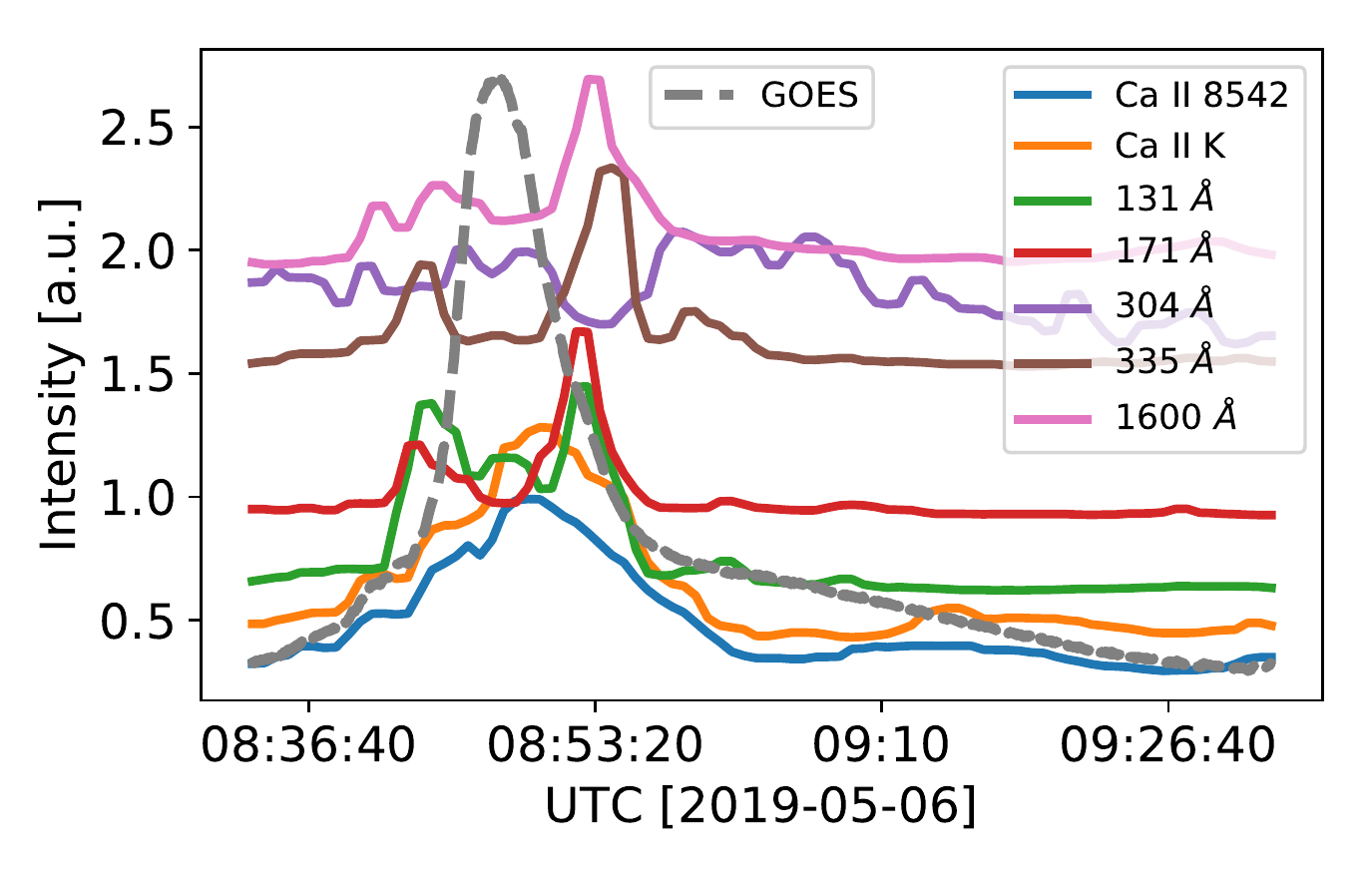}
\caption{\footnotesize Temporal evolution of intensity observed with the AIA channels and the \ion{Ca}{ii}~K and \ion{Ca}{ii}~8542~\AA\ lines for the pixels located at the points P1, P2, and P3 shown in Fig. \ref{fig_overview}. The dashed gray lines refer to the GOES X-ray curve. The profiles are arbitrarily shifted along the y-axis.}
\label{ligh_curve}
\end{figure*}

Figure \ref{ligh_curve} shows the temporal evolution of the observed intensity profiles at selected pixels. During the flaring time, we observed a strong intensity enhancement in all channels of the AIA, \ion{Ca}{II}~K, and \ion{Ca}{II}~8542~\AA\ lines.
The temporal evolution of X-rays observed by GOES is similar to the flare ribbon intensity observed in the \ion{Ca}{II} lines, which suggests that the intensity emission observed in the chromospheric lines might be mainly due to the heating of the chromosphere.

\section{Inversion of spectropolarimetric data}
\label{sec-inversion}
Generally, the magnetic field vector and other thermodynamic parameters of the solar atmosphere are inferred from the interpretation of the observed Stokes profiles. Nowadays, there are several available methods to do this, such as the  WFA (\citealt{2004ASSL..307.....L}) and the inversion of Stokes profiles under different conditions. Although non-LTE inversions are more accurate than the WFA, they are computationally more expensive. Therefore, in order to roughly infer the magnetic field at the photospheric and chromospheric levels for the full FOV, we employed a Milne-Eddington inversion code and the WFA, respectively, as explained in the following section.

\subsection{Milne-Eddington and weak-field approximation}
To infer the magnetic field in the photosphere (using the \ion{Fe}{I} 6173 line), we employed the SPIN code \citep{2017SoPh..292..105Y}, which is based on the analytic solution of the radiative transfer equation for polarized radiation in a Milne-Eddington model atmosphere \citep{1977SoPh...55...47A,2003isp..book.....D}. During the inversion of all the \ion{Fe}{i} scans, we took into account the instrumental profile of the CRISP instrument.

The observed linear and circular polarization signal strength in the \ion{Fe}{i} line was sufficient to infer the magnetic field vector in the photosphere. However,  in the chromosphere the linear polarization in the \ion{Ca}{ii}~8542~\AA\ line was not sufficient to make a fair estimate of the magnetic field vector in the FOV. On the other hand, the strength in the circular polarization was sufficient to estimate the LOS magnetic field (B$_{\rm LOS}$) using the WFA, which assumes that the Zeeman splitting is smaller than the Doppler broadening of the line. To this aim, we employed a spatially regularized WFA method\footnote{\url{https://github.com/morosinroberta/spatial\_WFA}}, which is based on the imposition of the Tikhonov regularization \citep{Tikhonov77}. More details regarding this algorithm are given in \cite{2020arXiv200614487M}. 

\subsection{Multiline inversion}
\label{sec_stic}
We employed the parallel non-LTE STockholm Inversion Code\footnote{\url{https://github.com/jaimedelacruz/stic}} (STiC; \citealt{delaCruz2016}; \citealt{delaCruz2019_STiC}) to simultaneously infer the atmospheric stratification of temperature, velocities, and magnetic field from the \ion{Fe}{i} 6173~\AA,\, \ion{Ca}{ii}~K, and \ion{Ca}{ii}~8542~\AA\ spectral lines. The STiC inversion code is built around a modified version of the RH code \citep{Uitenbroek2001} in order to derive the atomic populations by assuming statistical equilibrium and a plane-parallel geometry. The equation of state is borrowed from the Spectroscopy Made Easy (SME) computer code described in \cite{2017A&A...597A..16P}. The radiative transport equation is solved using cubic Bezier solvers \citep{2013ApJ...764...33D}.

For the inversion of the Stokes profiles, we considered the \ion{Ca}{II}~8542~\AA\ line in non-LTE conditions, under the assumption of complete frequency redistribution, while the \ion{Ca}{ii}~K line was synthesized in non-LTE conditions with partial redistribution effects of scattered photons. We used the fast approximation proposed by \cite{2012A&A...543A.109L}. The \ion{Fe}{I} 6173~\AA\ line was treated under the assumption of LTE conditions. We inverted all the Stokes parameters in the \ion{Fe}{I}~6173~\AA\ and \ion{Ca}{II}~8542~\AA\ lines, but only Stokes $I$ in the \ion{Ca}{ii}~K line.

In order to construct a good initial model for the inversion process, we first performed inversions of a few sets of pixels, mainly located near the flare ribbons, using three different cycles (see Table \ref{table_node}). In the first cycle we considered an atmospheric model derived from the FAL-C model \citep{Fontenla1993} by interpolation to 63 depth points from $\log\tau_{500} = -7.8$ to $+1.0$. The outputs from this first inversion were then used to train a neural network to retrieve the mapping between the Stokes parameters and the physical quantities of the output atmospheres. Then we used the new model atmosphere retrieved from the neural network as an initial guess model for the final inversion. Since the neural network outputs were already close to the final solution, the successive inversion process benefited from it in two different ways: Local minima were successfully avoided, and the number of iterations was significantly reduced, thus speeding up the whole inversion process and providing more reliable results. A similar approach making use of a neural network in order to produce an initial guess model was also performed by \cite{2020A&A...637A...1K}.

Since the inversion of all the frames in the whole FOV using the STiC code is computationally very expensive, only a few small patches of 5$\times$5 pixels highlighted in Fig. \ref{fig_overview} were selected for the inversion. 
To reduce the probability of falling in the local minima, we ran the inversion ten times for the third cycle, slightly modifying the outputs in each run. 

\begin{table}[]
\centering
\footnotesize
\caption{Number of nodes used for the temperature, LOS velocity (V$_{\rm LOS}$), turbulent velocity (V$_{\rm turb}$), LOS magnetic field (B$_{\parallel}$), horizontal magnetic field (B$_{\perp}$), and azimuth ($\phi$) during each cycle of the inversion.}
\begin{tabular}{|l|c|c|c|}
\hline

\label{table_node}

Physical parameters     & Cycle 1 & Cycle 2 & Cycle 3 \\
\hline
T                       & 7       & 9       & 10      \\
V$_{\rm LOS}$           & 2       & 5       & 7       \\
V$_{\rm turb}$          & 1       & 3       & 5       \\
B$_{\parallel}$         & 1       & 2       & 3       \\
B$_{\perp}$             & 1       & 2       & 3       \\
$\phi$                  & 1       & 1       & 2       \\  
\hline
\end{tabular}
\end{table}

\section{Results}
\label{sec_results}

\begin{figure*}[!t]
\centering
\includegraphics[width=.54\linewidth]{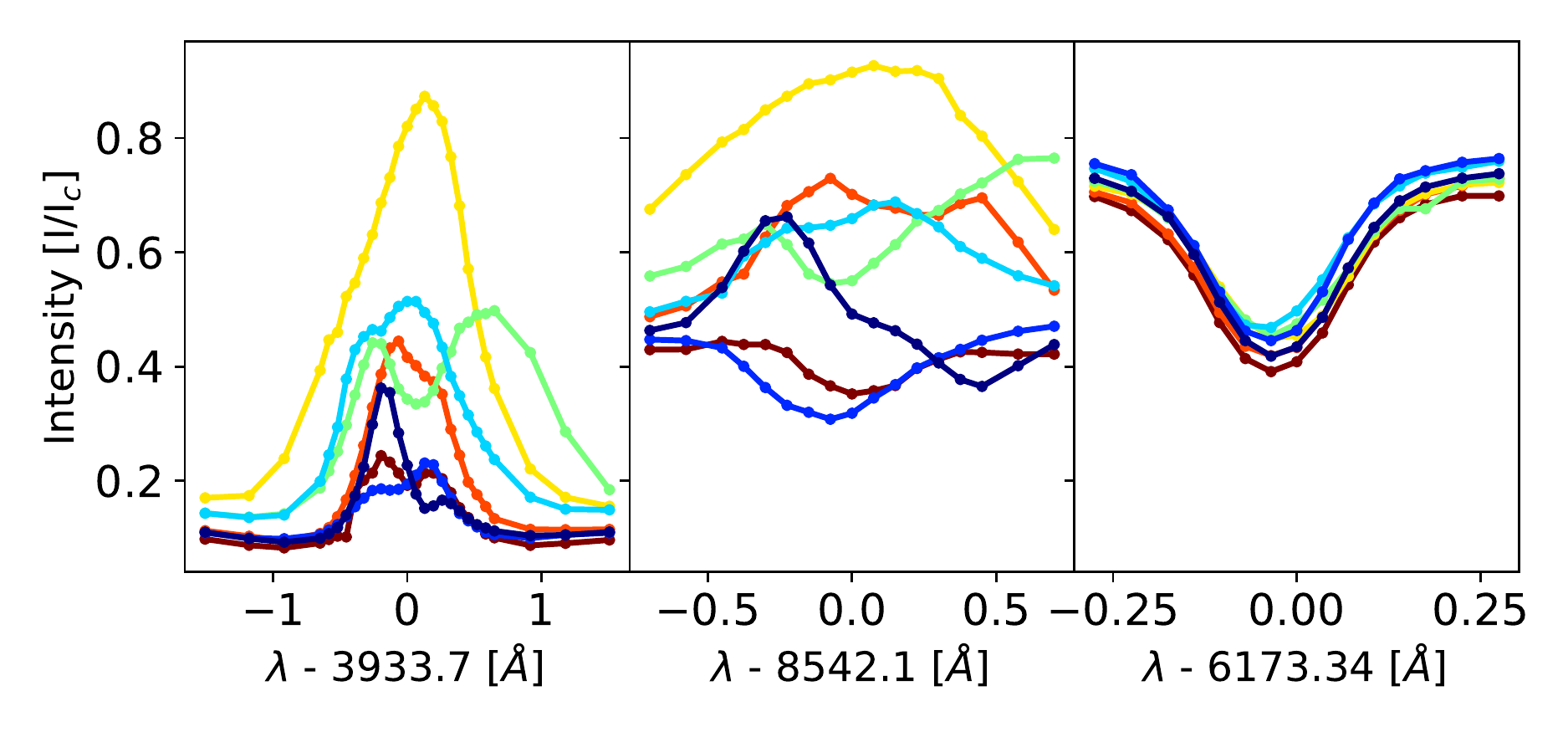} 
\includegraphics[width=.43\linewidth]{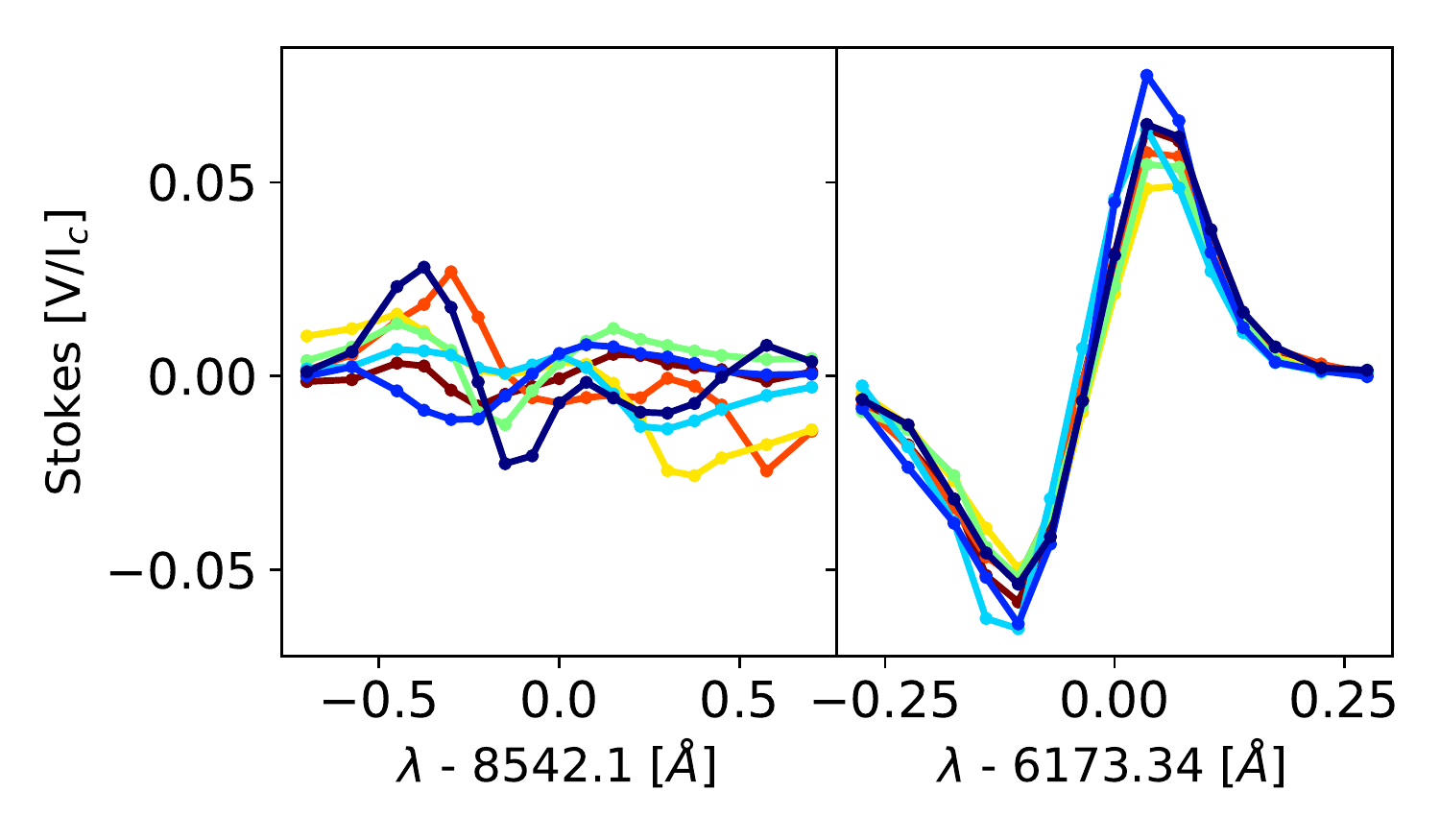}
\includegraphics[width=.54\linewidth]{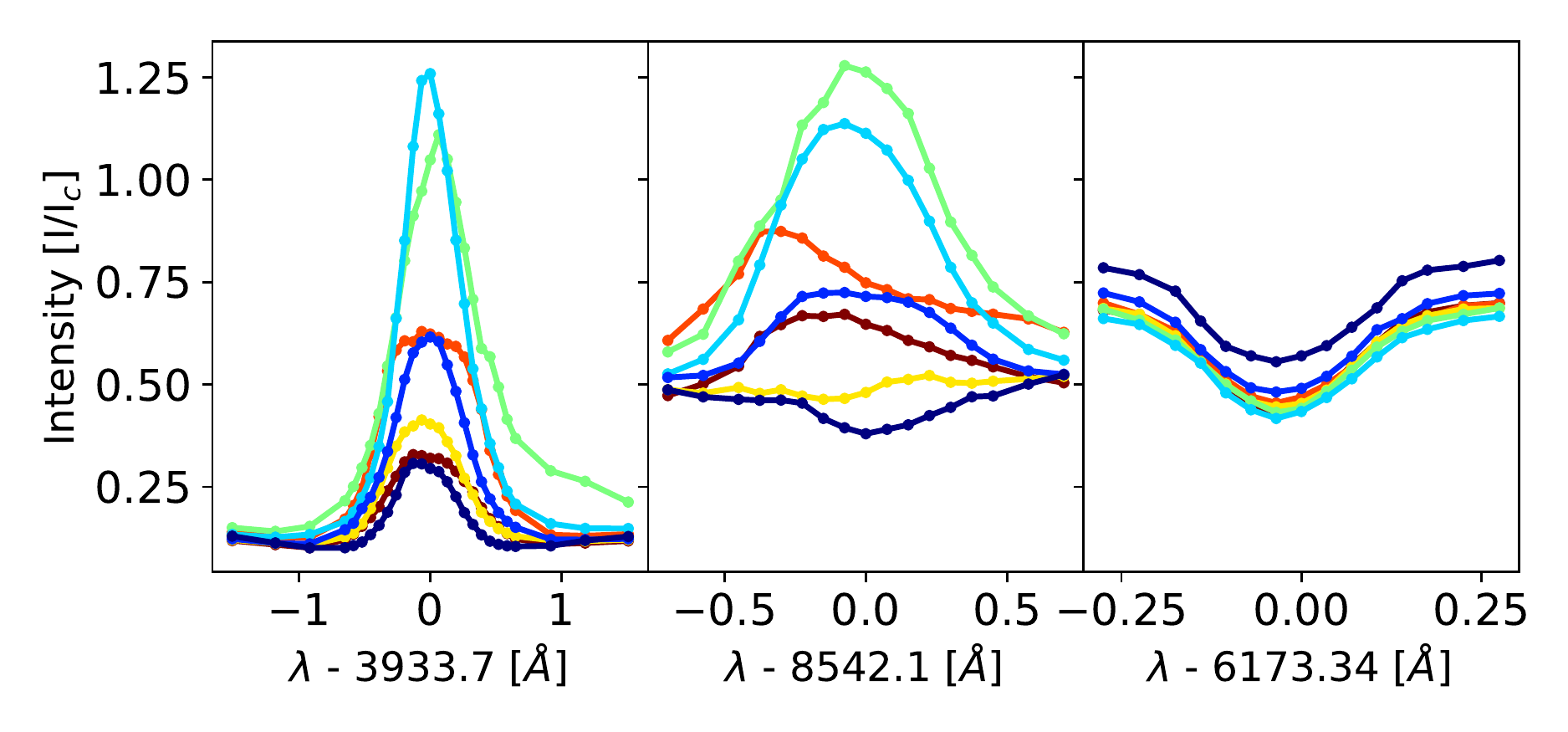} 
\includegraphics[width=.43\linewidth]{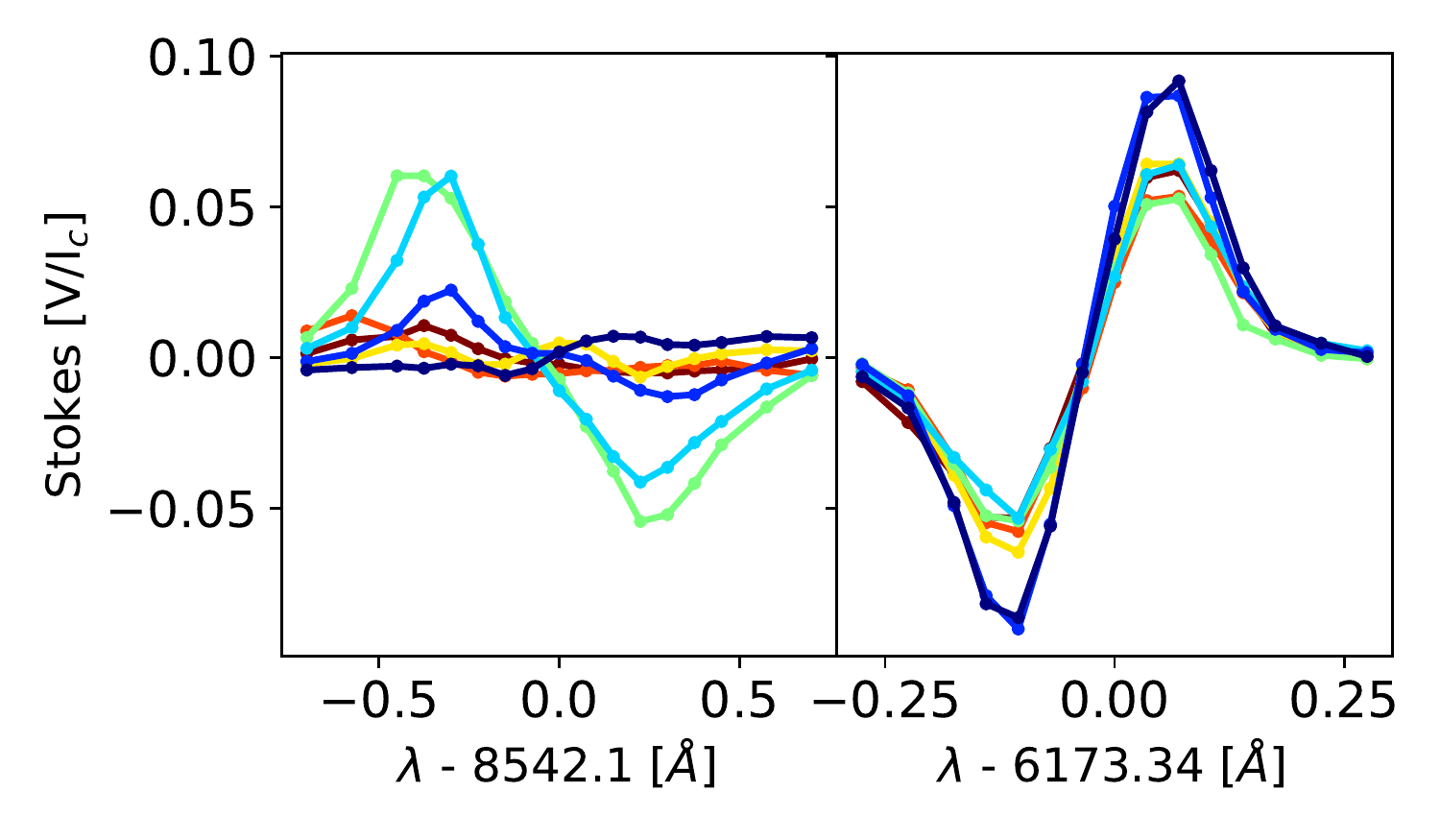}
\includegraphics[width=.54\linewidth]{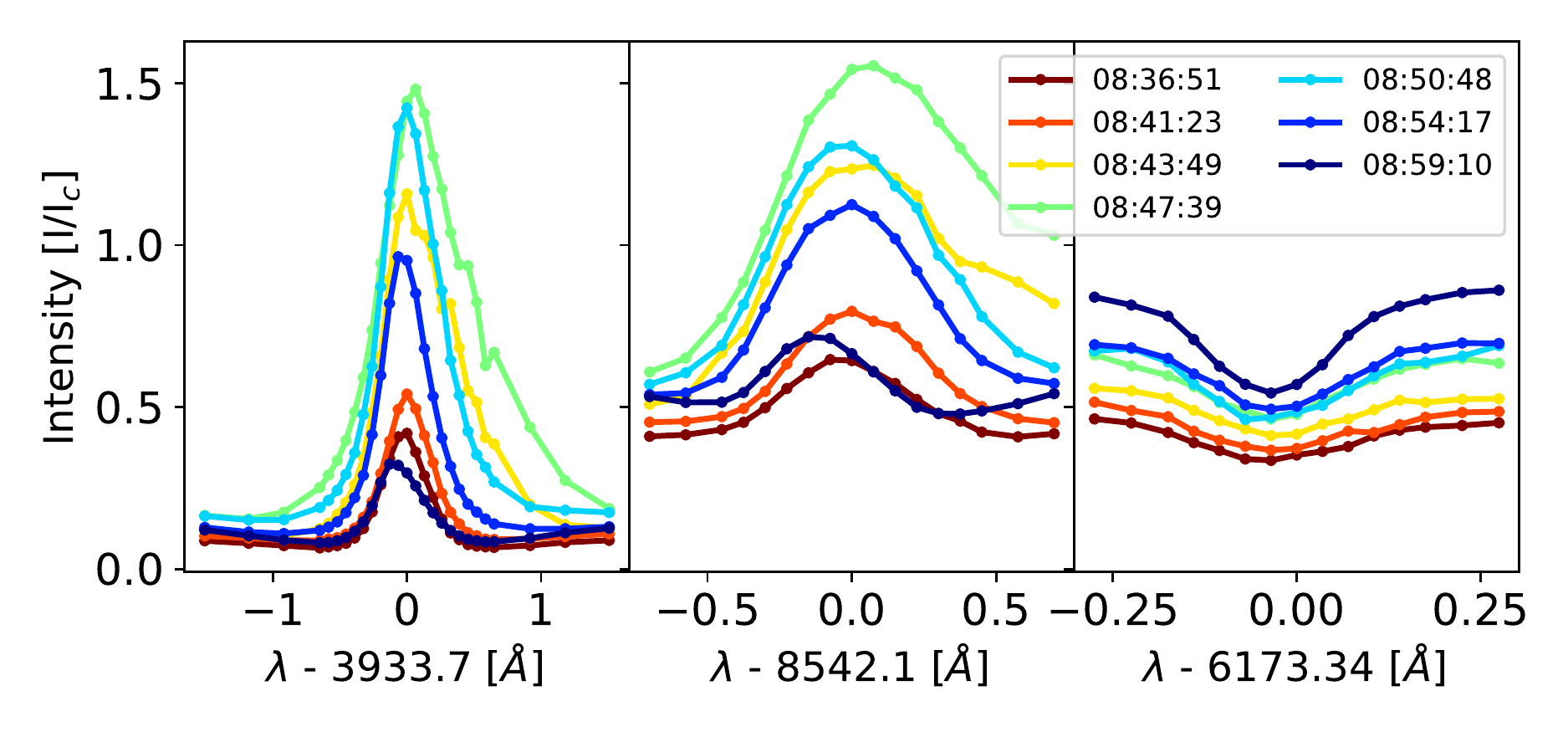} 
\includegraphics[width=.43\linewidth]{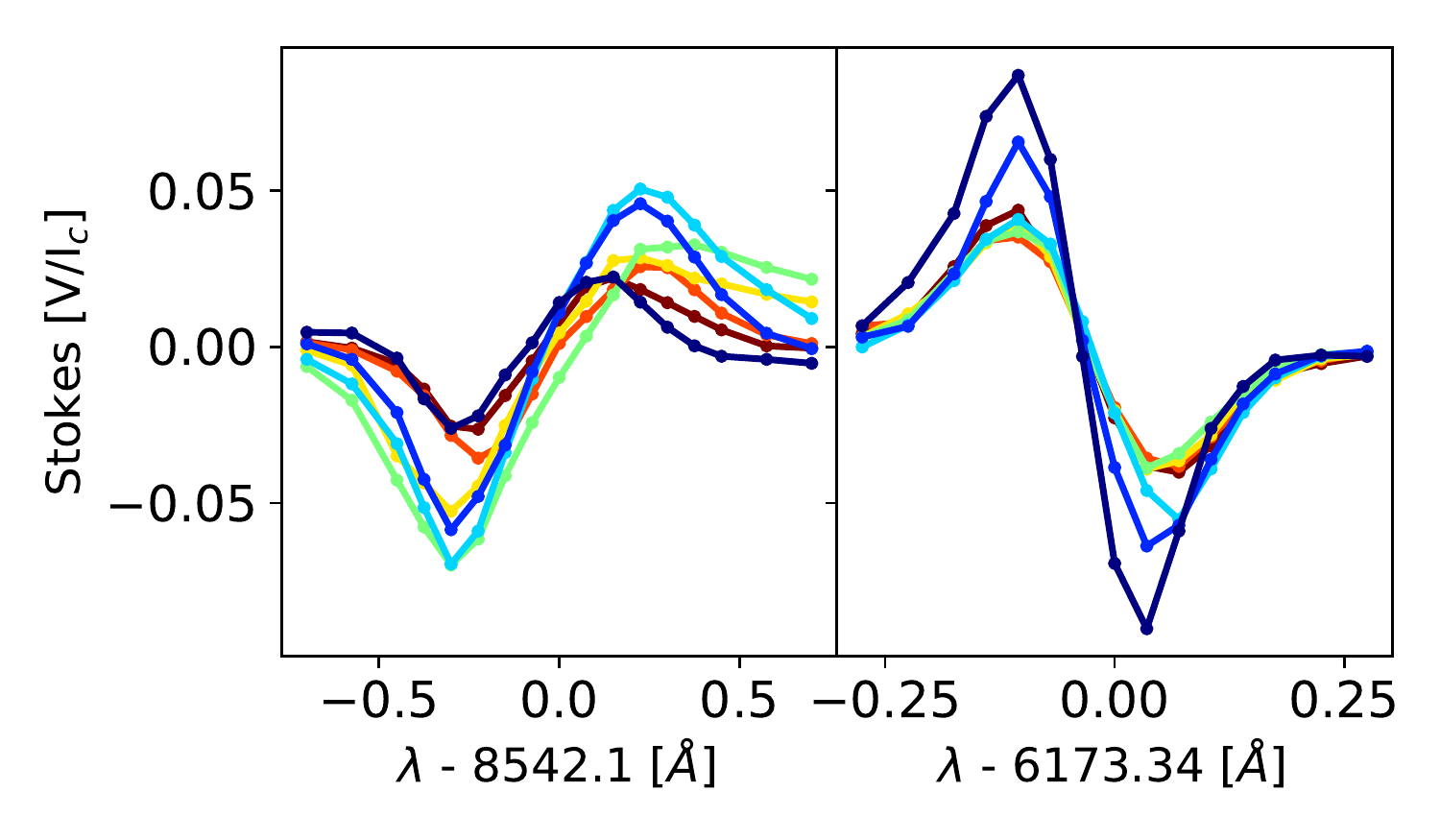}

\caption{ Temporal evolution of the observed Stokes~$I$ (\ion{Ca}{ii}~K, \ion{Ca}{ii}~8542~\AA,\ and \ion{Fe}{i}~6173~\AA\ lines) and Stokes~$V$ profiles (\ion{Ca}{ii}~8542~\AA\ and \ion{Fe}{i}~6173~\AA\ lines) around the flare start, peak, and end times at the locations of the points P1 (top panels), P2 (middle panels), and P3 (bottom panels) shown in Fig.~\ref{fig_overview}. Different colors depict different times.}
\label{intensity_profiles}
\end{figure*}


\subsection{Temporal evolution of Stokes profiles}
Figure \ref{intensity_profiles} shows the temporal evolution of Stokes profiles near the flare ribbon position indicated in Fig.~\ref{fig_overview}. The observed Stokes profiles are complex in nature. For instance, the Stokes~$I$ profiles of \ion{Ca}{II} lines are broad, asymmetric, and have a single emission peak around the flare peak time.
They show an increase in the intensity as the flare starts, become the strongest near the flare peak time, and decrease gradually during the decay phase of the flare. 
For example, the intensity profile of a pixel located at the flare footpoint shows an increase by a factor of $\sim~$8~(4) relative to the same pre-flare pixel in the \ion{Ca}{II}~K~(\ion{Ca}{II} 8542) intensity at the flare peak time.

In contrast to the \ion{Ca}{ii}~8542~\AA\ line, the \ion{Ca}{ii}~K line shows a fast response to change in the intensity profiles. This can be related to the \ion{Ca}{ii}~K line formation height, which forms relatively higher in the atmosphere compared to the \ion{Ca}{ii}~8542~\AA\ line \citep{2019A&A...631A..33B}. 
The intensity enhancement in the line core of \ion{Ca}{ii}~8542 has also been reported in flare observations by different authors  \citep{2012ApJ...748..138K,Kuridze2018}. Around the flare peak time the intensity profile in both \ion{Ca}{ii} lines shows drastic change and becomes more asymmetric relative to the start and end times of the flare. 
Unlike what we observed in the chromospheric spectral lines, no significant intensity enhancement is observed in the photospheric \ion{Fe}{I}~6173~\AA\ line, as displayed in Fig.~\ref{intensity_profiles}, which could be due to the lesser influence of the  C-class flare.

The temporal evolution of the Stokes~$V$ profiles at the flare ribbon locations shows amplitude changes in both the \ion{Fe}{I} line and in the \ion{Ca}{ii}~8542~\AA\  lines. The polarity above the flare ribbons is the same at the two corresponding depth ranges mapped by the spectral lines, as expected. However, we observe the reversal in the chromospheric Stokes~$V$ profiles. This change is induced by a change in the gradient of the source function as a function of depth, which makes the core of Stokes~$I$ appear in emission and causes a reversal in the polarity in the Stokes~$V$ profiles. This behavior is also present in umbral flashes \citep[e.g.,][]{2013A&A...556A.115D}.

Similar reversal profiles were also reported by \cite{2012ApJ...748..138K} in a C-class flare. It is clear from Fig.~\ref{intensity_profiles} that there is an enhancement in the chromospheric Stokes~$V$ signal around the flare peak time, and after the flare the signal in Stokes~$V$ decreases. Moreover, highly asymmetric Stokes~$I$~and~$V$ profiles indicate the presence of a strong gradient in the LOS velocity. Noticeably, during the flare the profiles exhibit changes on a timescale of a few seconds, which is also noticed in flare simulations \citep{2019NatAs...3..160C, 2020ApJ...895....6G}.

In our observations the acquisition time needed to scan the \ion{Ca}{ii}~K line is $\sim\,$15~s, which implies that part of the asymmetric shape of the profiles could be due to the unresolved fast temporal changes of the profiles. 
The asymmetry in the profiles can also be introduced by the presence of multi-components of model parameters, which can appear in dynamic features such as a flare. Even though the changes in the profiles of the \ion{Ca}{ii}~8542~\AA\  and \ion{Ca}{ii}~K lines are similar for most of the pixels, they are not exactly co-temporal. The CRISP and CHROMIS instruments recorded data quasi-simultaneously but not at the exact same time. This difference can also alter the inference of the model parameters obtained from the inversion code; this is an instrumental limitation inherent to imaging spectropolarimetry.

\begin{figure}[!t]
\centering
\includegraphics[width=1\linewidth]{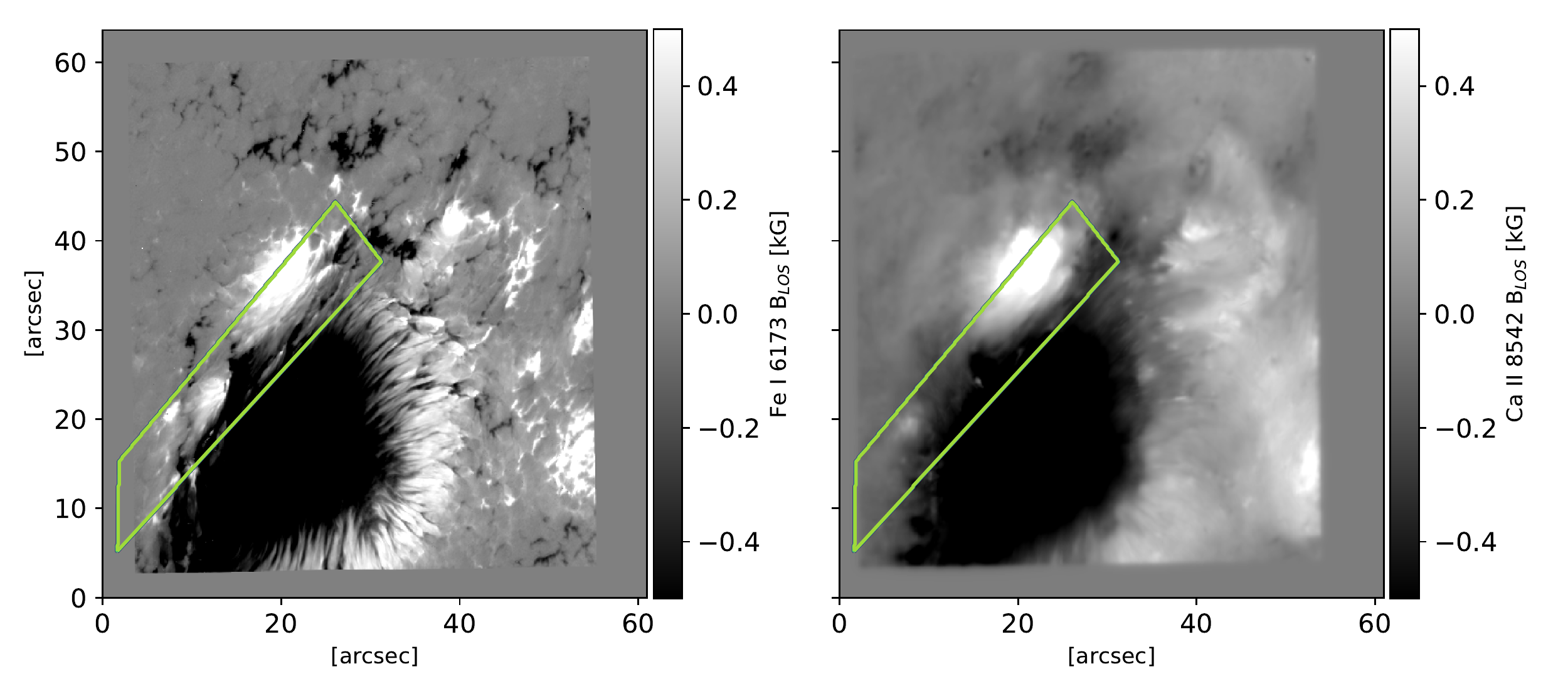}
\includegraphics[width=.49\linewidth]{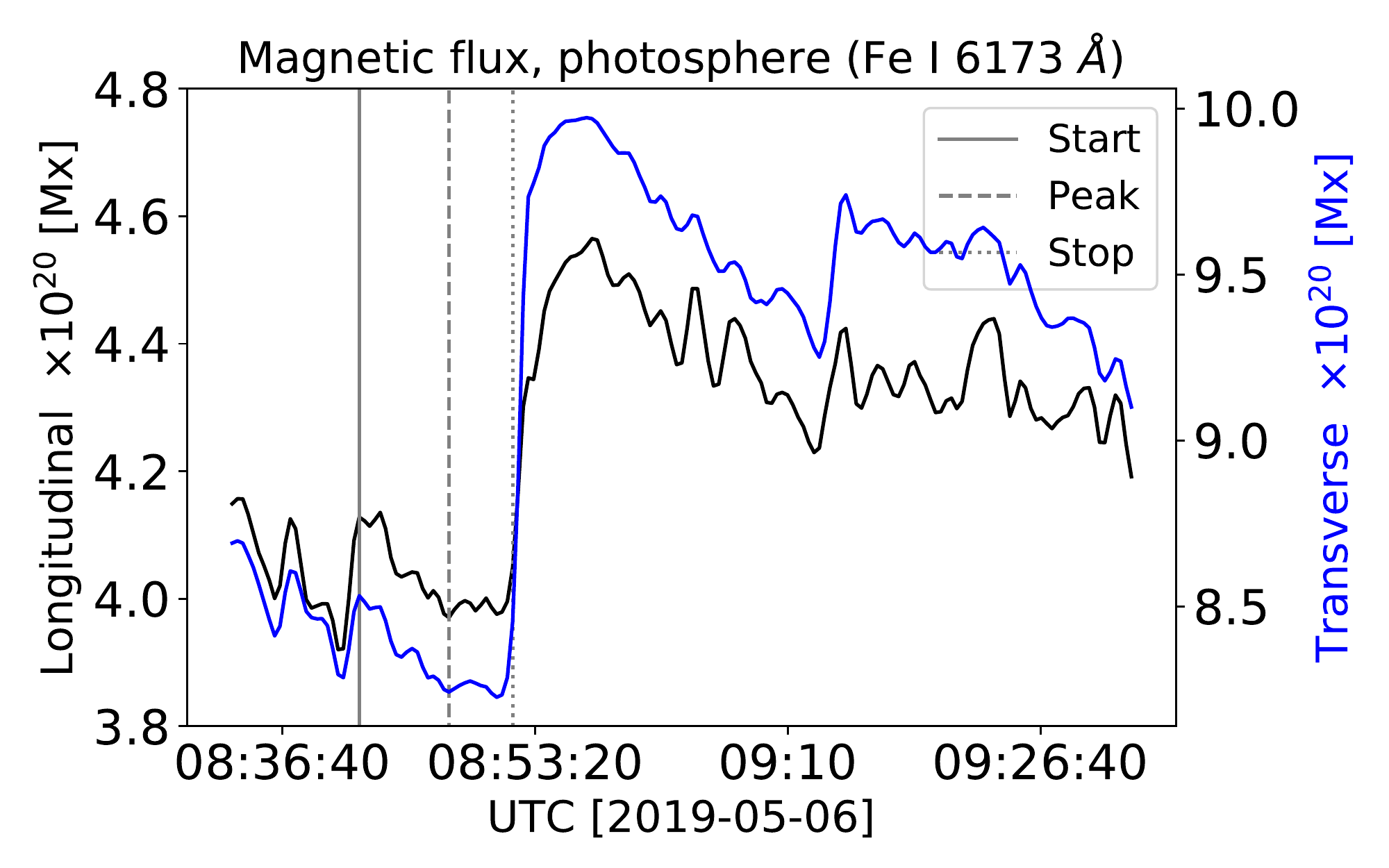}
\includegraphics[width=.49\linewidth]{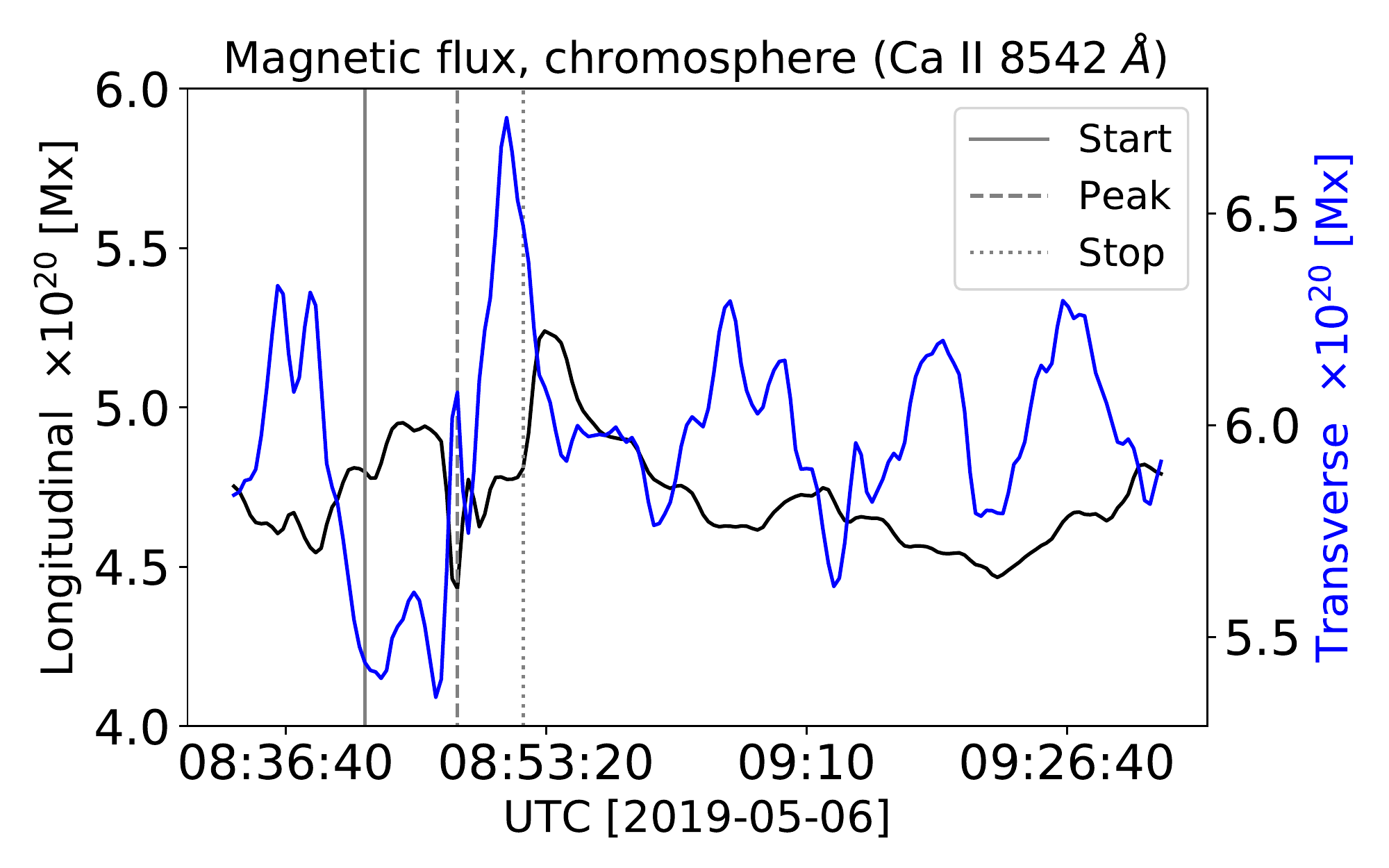}
\includegraphics[width=.49\linewidth]{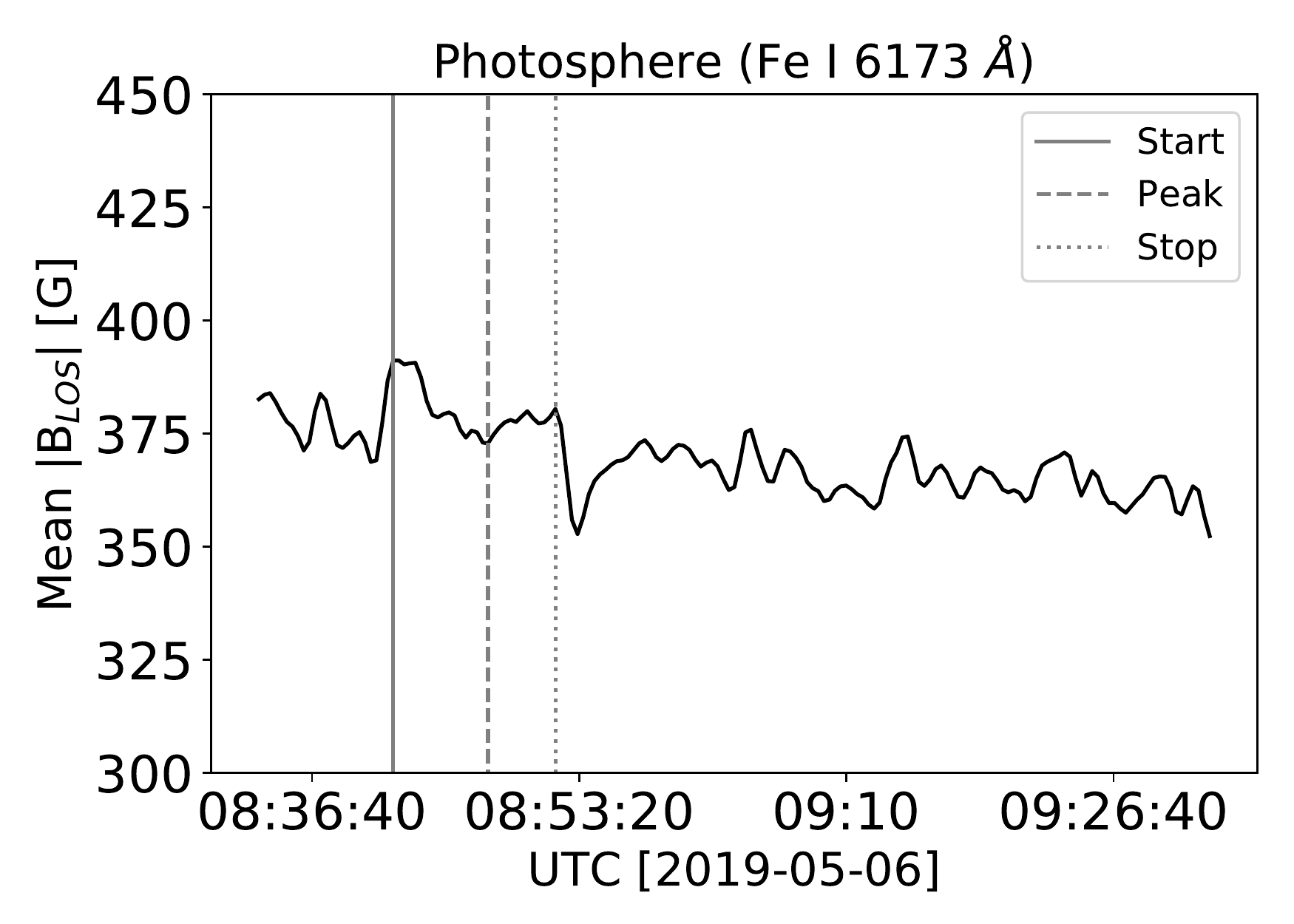}
\includegraphics[width=.49\linewidth]{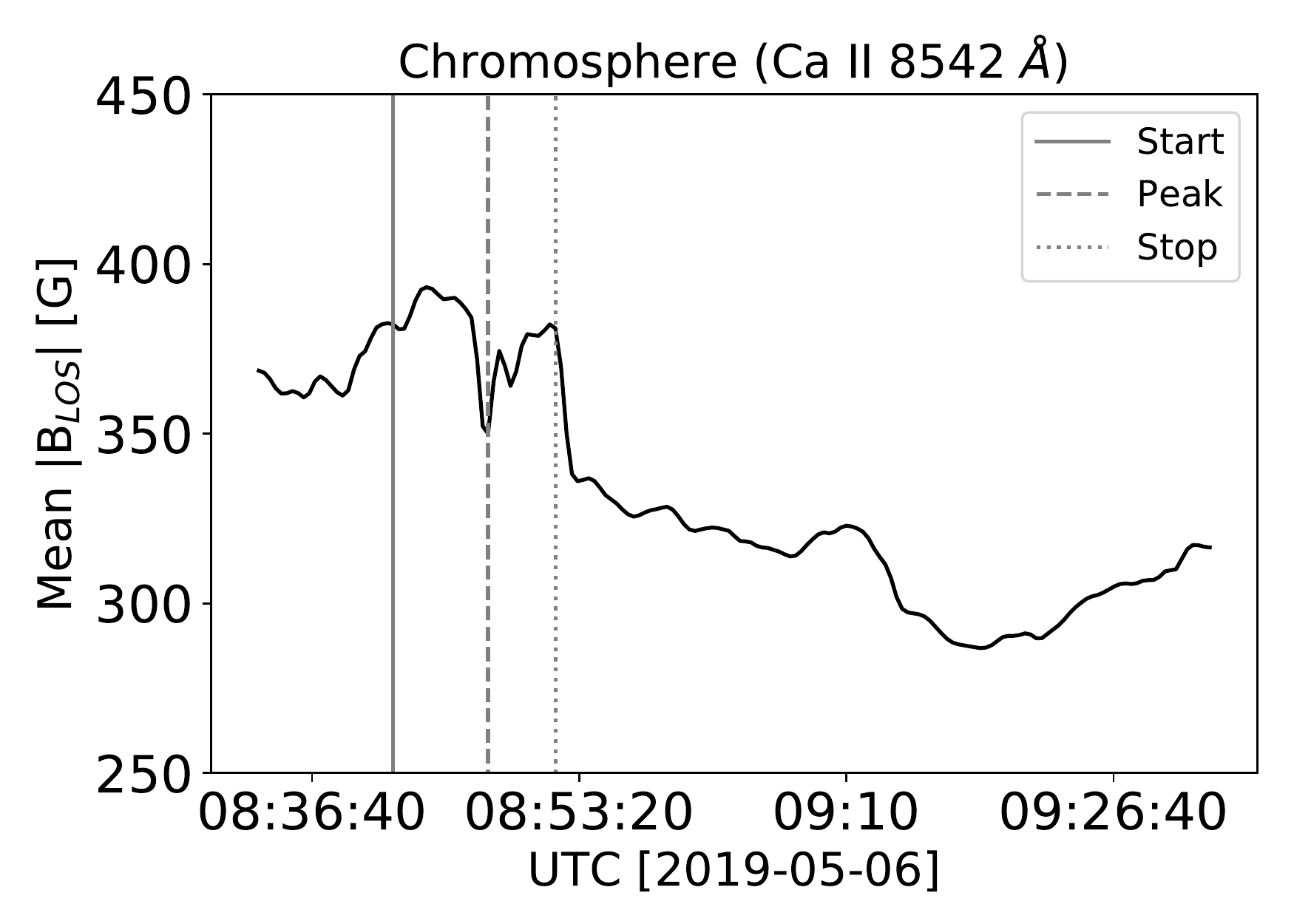}
\includegraphics[width=1\linewidth]{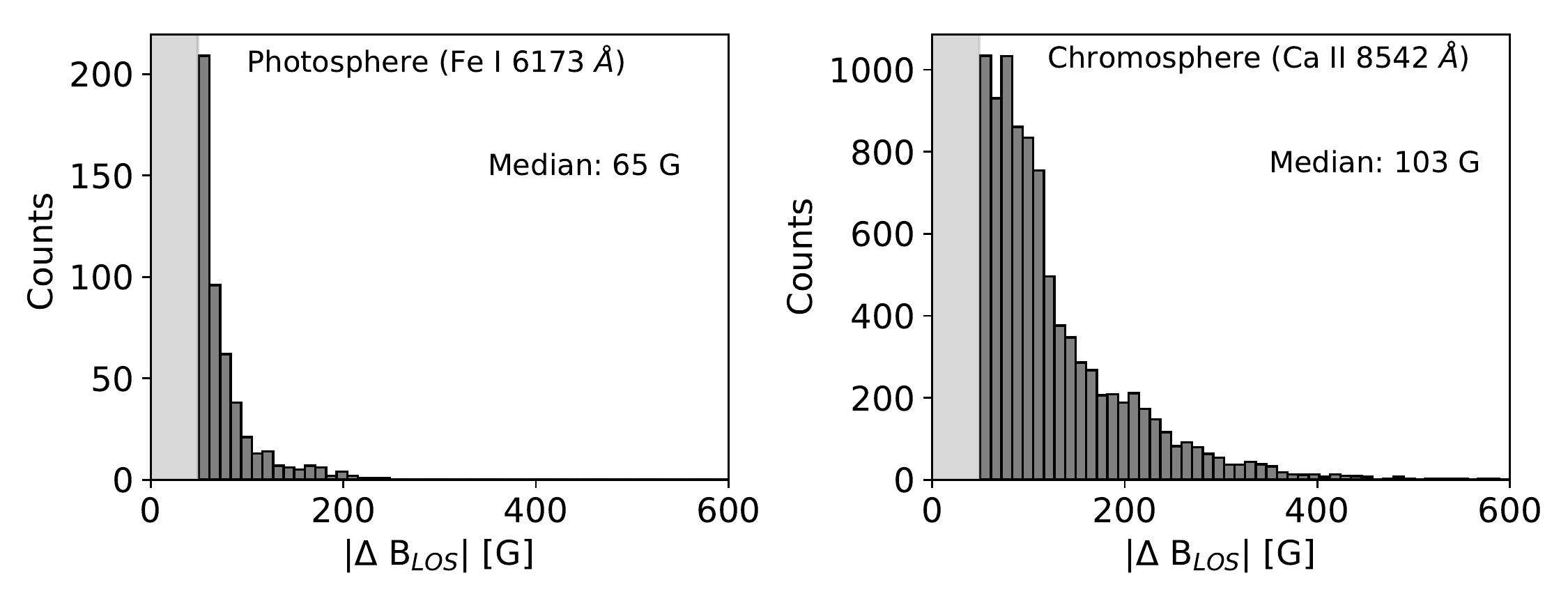}

\caption{\footnotesize Retrieved magnetic field in the photosphere (left column) and chromosphere (right column) during the flare. Top row: Photospheric magnetogram retrieved from a Milne-Eddington inversion of the \ion{Fe}{i} line (left), and chromospheric magnetogram obtained from the WFA for the \ion{Ca}{ii}~8542~\AA\ line around the flare peak time (08:46 UT; right). Second row: Temporal evolution of the mean longitudinal and transverse magnetic flux in the green box. Third row: Mean LOS magnetic field evaluated for the pixels located in the green box. Bottom row: Histograms of the LOS magnetic field change ($>$50~G) observed in the photosphere and the chromosphere at the flare peak time. }
\label{flux_change_phot_chrom}
\end{figure}

\subsection{Magnetic field and flux in the photosphere and chromosphere}

In this section we present the temporal evolution of the magnetic field parameters inferred in the photosphere and in the chromosphere during the flare. As the linear polarization signal in the \ion{Ca}{ii} 8542~\AA\ line is not sufficiently high to infer the magnetic vector in the chromosphere, we only analyzed the B$_{\rm LOS}$ in the chromosphere and photosphere, which was estimated using the SPIN code and the WFA, as discussed in Sect. \ref{sec-inversion}.

The top panels of Fig.~\ref{flux_change_phot_chrom} display the B$_{\rm LOS}$ maps during the flare peak time in the photosphere and in the chromosphere. They show that the chromospheric B$_{\rm LOS}$ decreases and becomes smoother compared to the photospheric B$_{\rm LOS}$. 
During the flare, the longitudinal and transverse magnetic fields in the photosphere and in the chromosphere exhibit changes, mainly near the location of the polarity inversion line  (PIL) and the flare ribbons \citep{2010ApJ...724.1218P, 2017ApJ...834...26K, 2018ApJ...852...25C}. Therefore, for further analysis we selected a region, highlighted in green in the top panel of Fig.~\ref{flux_change_phot_chrom}, that encompasses both the polarity inversion line and flare ribbons.

In the photosphere, we found that the mean longitudinal and transverse magnetic fluxes, within the green box, changed at 08:53~UT, 6 minutes after the flare peak. It is not clear whether this change is due to the flare or to the emergence of flux in the selected region. More quantitatively, the longitudinal and transverse magnetic fluxes show relative changes of $\sim~$22\% and 10\%, respectively. Similarly, in the chromosphere, changes in the longitudinal and transverse magnetic fluxes during the flaring time are also visible (see the second row of Fig. \ref{flux_change_phot_chrom}).
The changes started around 08:40~UT, and the initial configuration was partially and gradually restored after the flare. The estimated maximum relative changes before and after the flare in the longitudinal and transverse magnetic fluxes were $\sim\,$26\% and $\sim\,$6\%, respectively. These changes were significant during the flaring time, which could be related to the flare as it can change the magnetic field configuration in the upper solar atmosphere.

The mean photospheric LOS magnetic field in the selected region 
exhibited negligible changes before and after the flare, as displayed in the third row of Fig. \ref{flux_change_phot_chrom}. However, in the chromosphere, they showed abrupt changes during the flaring time. Around the flare start time, the B$_{\rm LOS}$ in the chromosphere increased slowly, and after the flare it decreased gradually to $\sim$300~G.

We also analyzed the change in B$_{\rm LOS}$ in the selected region, pixel by pixel, in the photosphere and the chromosphere. However, we did not notice a step-wise change in B$_{\rm LOS,}$ as reported by different authors, neither in the photosphere \citep{2018ApJ...852...25C,2010ApJ...724.1218P}, nor in the chromosphere \citep{2017ApJ...834...26K}. Such a different behavior could be attributed to the less intense C-class fare.
Nevertheless, our results are consistent with \cite{2018ApJ...852...25C}, who reported that flares larger than the M1.6-class typically show step-wise changes in the photosphere. Furthermore, in their investigation, the smallest flare displaying changes in B$_{\rm LOS}$ was a C3-class flare, which is comparatively stronger than the flare analyzed in the present study.

The histograms in the bottom panels of Fig.~\ref{flux_change_phot_chrom} display the difference between the B$_{\rm LOS}$ at the flare peak and at the no-flare times in the photosphere and in the chromosphere, determined from the pixels located in the green box.
It shows that the chromospheric B$_{\rm LOS}$ changes by up to $\sim$~400~G (median $\sim103$~G), whereas the corresponding photospheric change stays below 200~G (median $\sim65$~G). Similar strong changes in the chromosphere were also reported by \cite{2017ApJ...834...26K}, but for an X1-class flare.

\subsection{Inversion of the flare footpoints}
The physical parameters, such as temperature, magnetic field, LOS velocity, and microturbulent velocity, were inferred by inverting three lines (\ion{Fe}{i}~6173~\AA, \ion{Ca}{ii}~K, and \ion{Ca}{ii}~8542~\AA) simultaneously. For this purpose, we employed the STiC code (see Sect. \ref{sec_stic}). The stratification of the physical parameters were obtained as a function of the logarithm of the optical depth scale at 500 nm ($\log\tau_{500}$).
The photospheric parameters were inferred from the \ion{Fe}{i}~6173~\AA\ line, whereas the lower and the middle chromospheric parameters were derived using the \ion{Ca}{ii}~K and the \ion{Ca}{ii}~8542~\AA\ lines. We calculated the sensitivity of the lines and found that the \ion{Ca}{ii}~K and \ion{Ca}{ii}~8542~\AA\ lines are less sensitive below the $\log\tau_{500}\sim-5$ (see Sect. \ref{sec_response}). Below that layer, uncertainties are thus expected to grow. Therefore, in this paper, we discuss the inferred parameters below $\log\tau_{500}\sim-5$ only.

\begin{figure}[!t]
\centering
\includegraphics[width=1\linewidth]{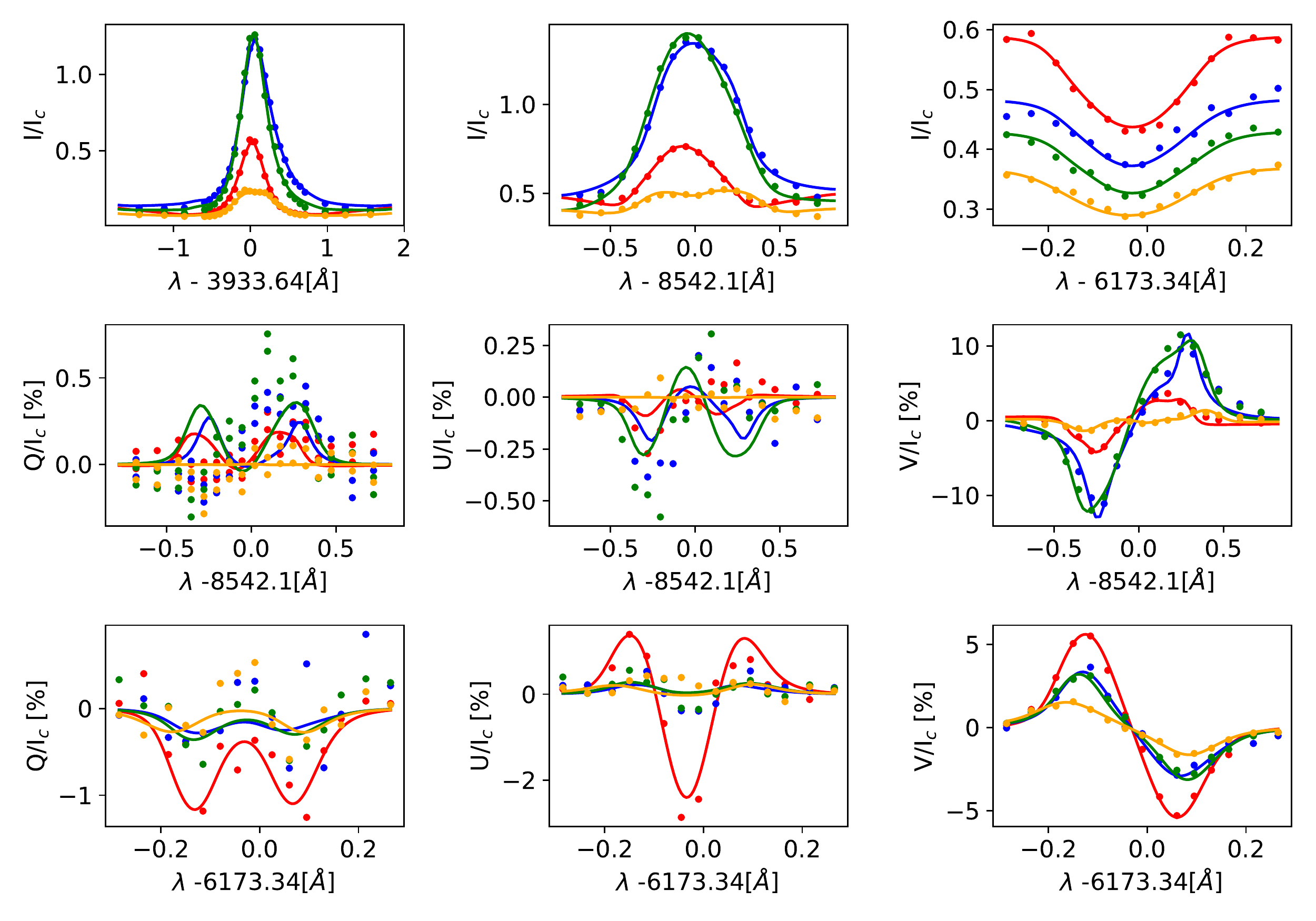}
\includegraphics[width=1\linewidth]{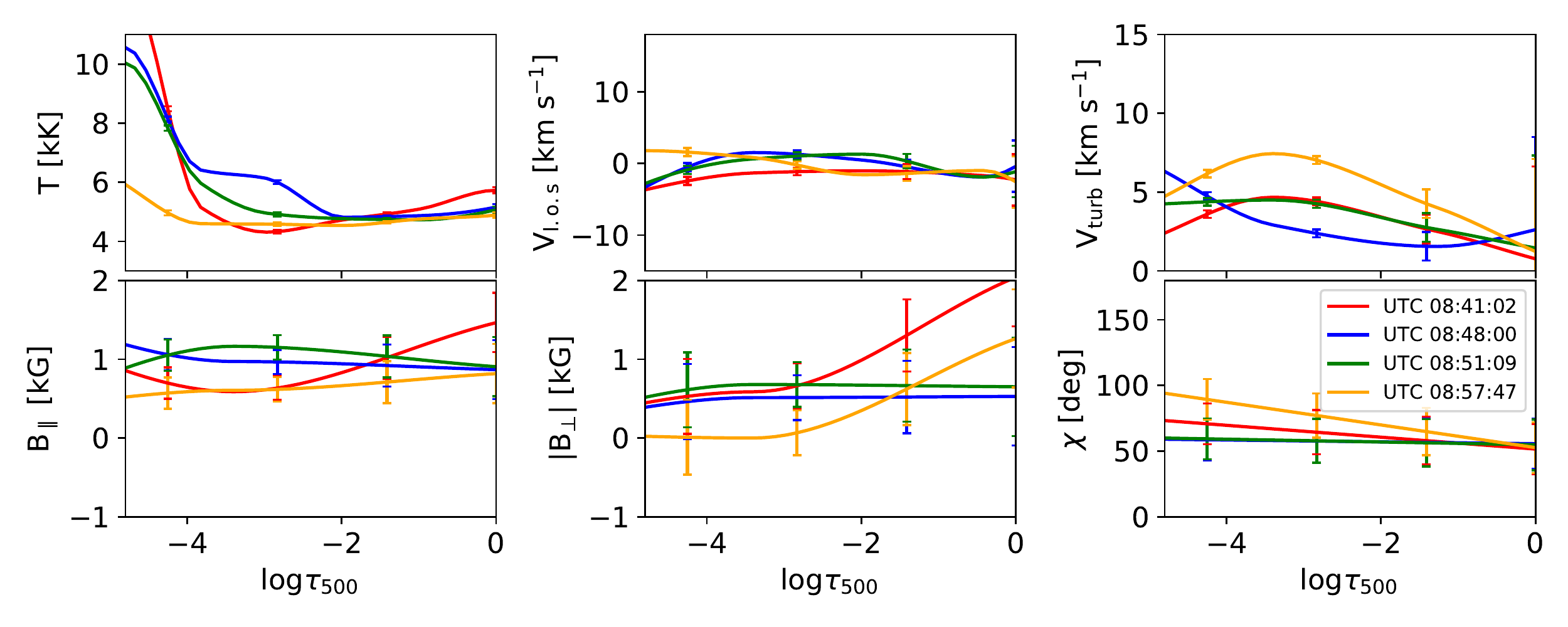}

\caption{\footnotesize Fitting of observed profiles and retrieved physical parameters. Top three rows: Observed (dotted lines) and best fits (solid lines) of the Stokes profiles. Bottom two rows: Inferred model atmosphere retrieved from the  inversion with the STiC code. Different colors indicate the time of the observations during a flare. The observed profiles are located at the P3 indicated in Fig.~\ref{fig_overview}. The vertical bars refer to the uncertainties in the inferred parameters at different node locations. }
\label{best_fit_330_190}
\end{figure}

\begin{figure}[!h]
\centering
\includegraphics[width=1\linewidth]{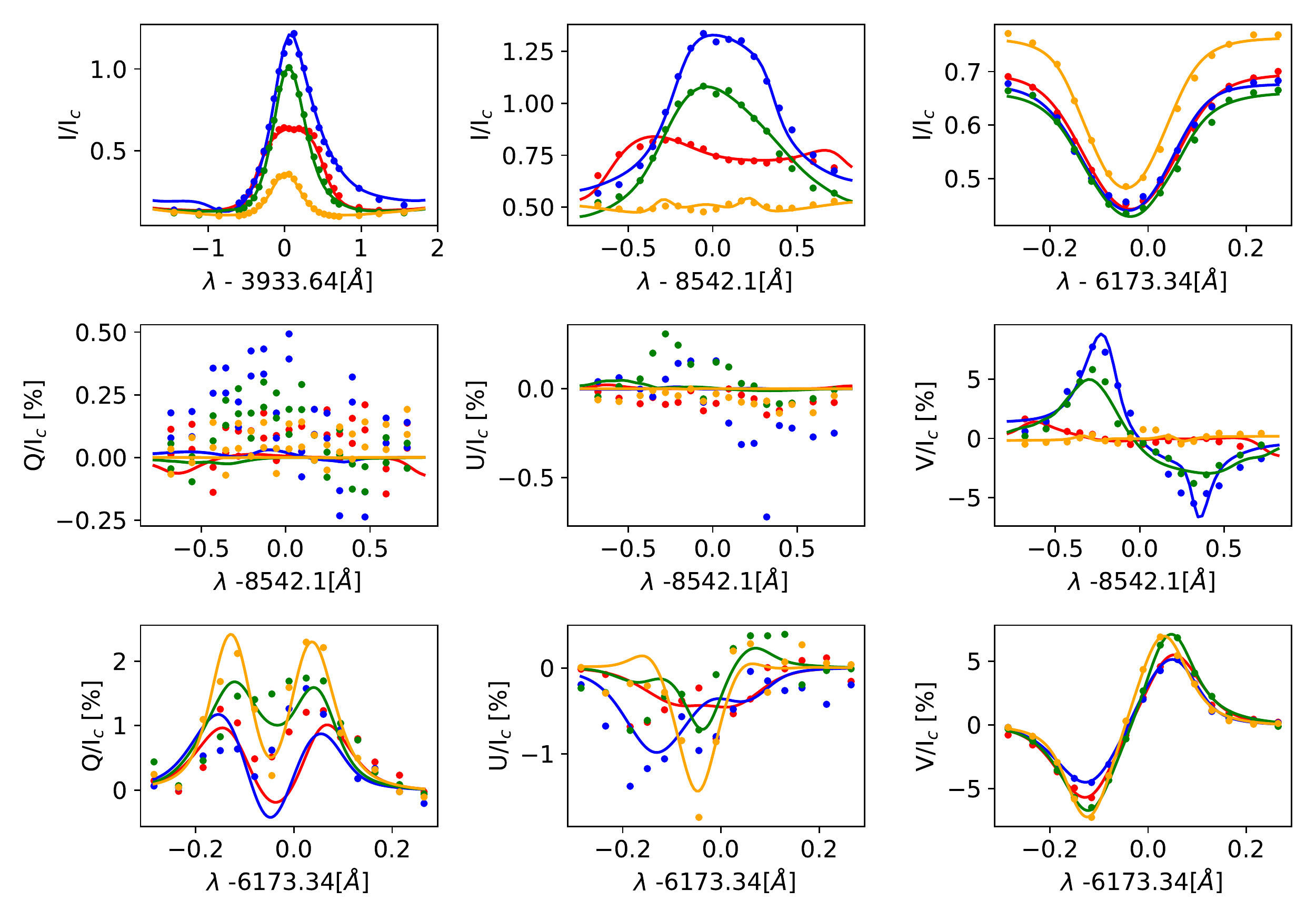}
\includegraphics[width=1\linewidth]{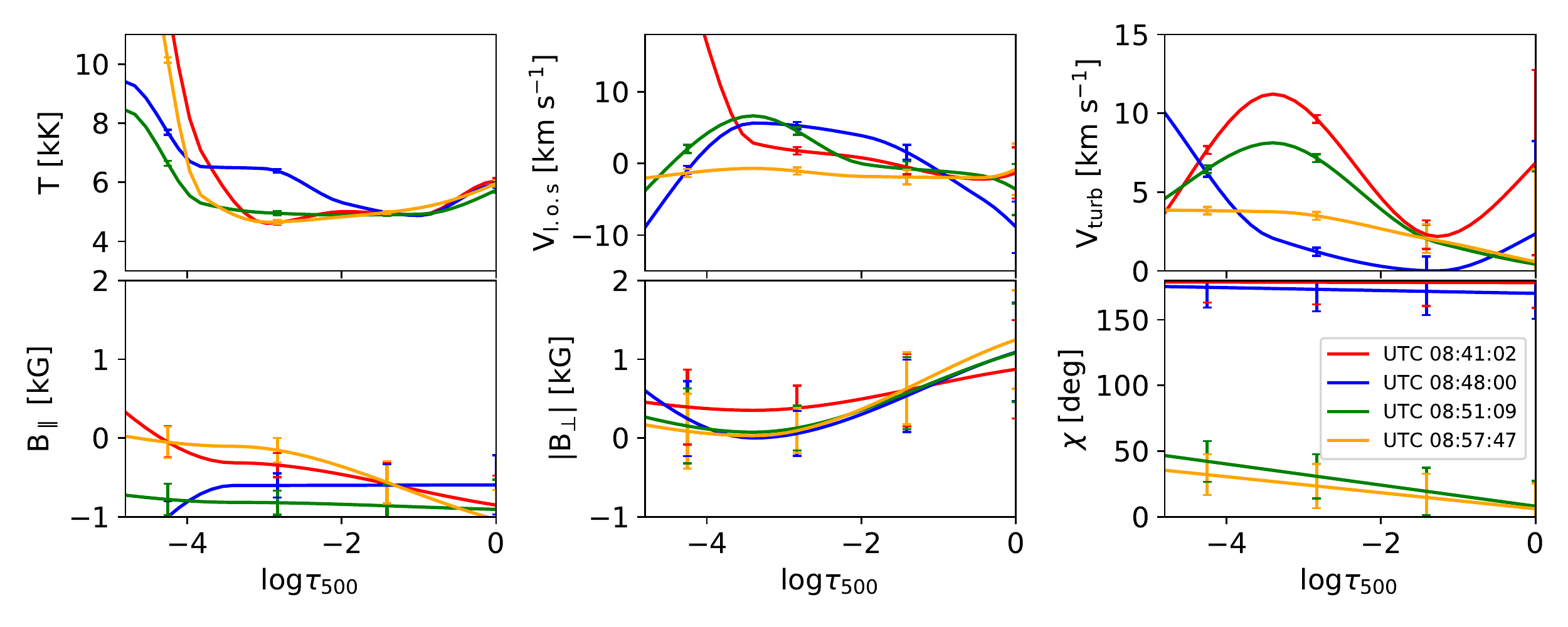}

\caption{\footnotesize Same as Fig. \ref{best_fit_330_190}, but for the pixels located at P2.}
\label{best_fit_220_180}
\end{figure}

\begin{figure}[!h]
\centering
\includegraphics[width=1\linewidth]{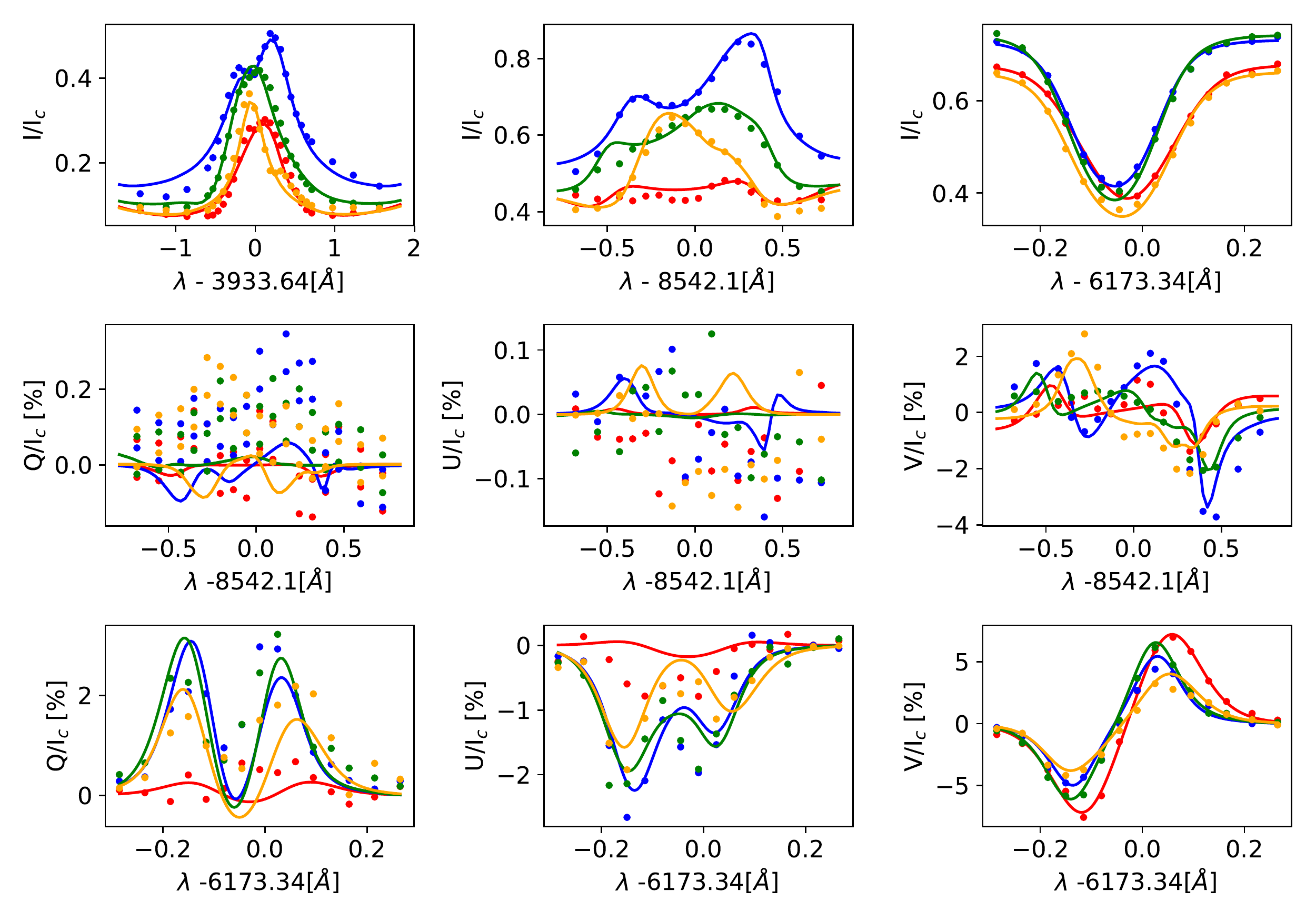}
\includegraphics[width=1\linewidth]{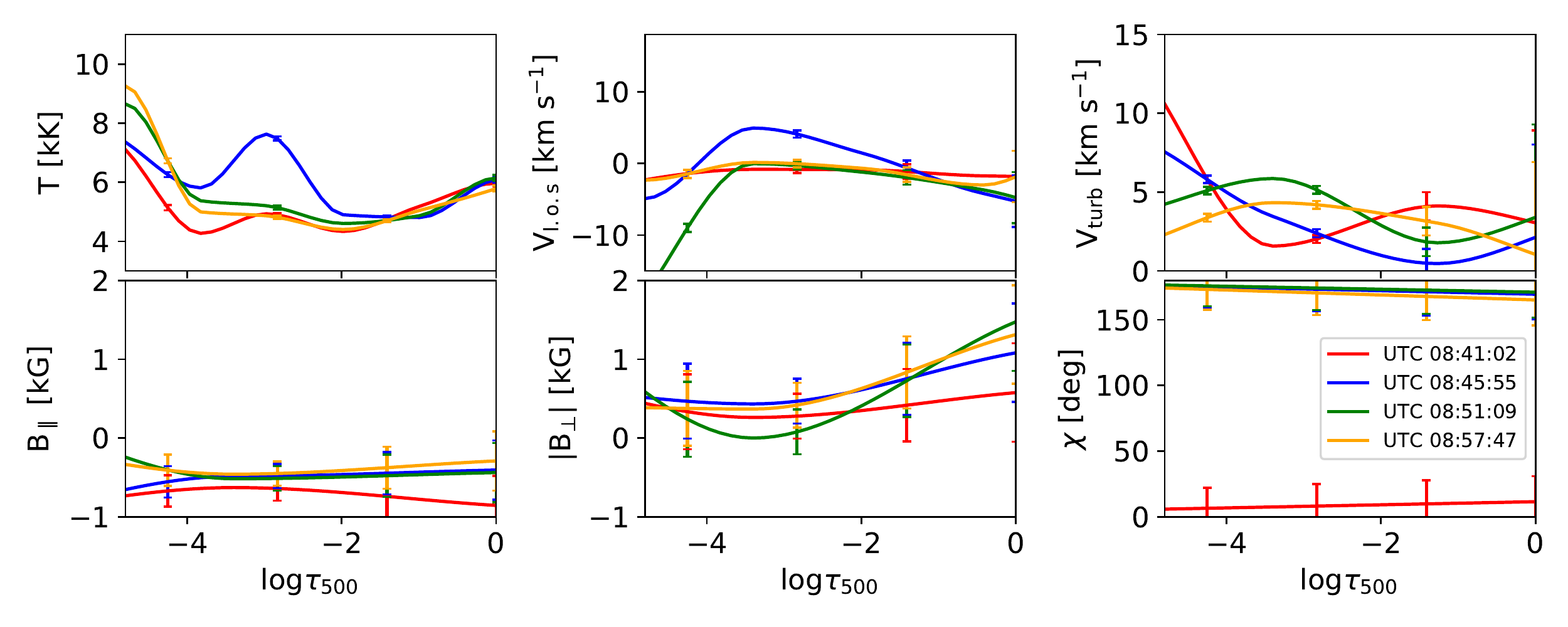}

\caption{\footnotesize Same as Fig. \ref{best_fit_330_190}, but for the pixels located at P1.}
\label{best_fit_185_135}
\end{figure}

Figures~\ref{best_fit_330_190}--\ref{best_fit_185_135} show some examples of observed and best-fit profiles for the selected pixels.  
They are displayed at different time steps of the flare, that is, near the start (red), peak (blue), and end (green) times. For comparison, we also show the inversion of Stokes profiles obtained $\sim$8 minutes after the flare end time (indicated with orange). The code clearly fails to fit the Stokes~$Q$ and $U$ profiles in the \ion{Ca}{ii}~8542~\AA\ line because of signals below the noise level. However, a reasonably good fit is achieved for the Stokes~$I$ and $V$ profiles.

The stratification of the inferred temperature, LOS velocity, microturbulent velocity, longitudinal and transverse magnetic fields, and azimuth angles is also shown with the same color coding in the bottom rows of Figs.~\ref{best_fit_330_190}--\ref{best_fit_185_135}. 
The inferred temperatures show that the atmosphere is heated up to $\sim$7.5~kK between $\log\tau_{500}\sim-4$ and $\log\tau_{500}\sim-2.5$ around the flare peak time.
Near the flare peak time, the retrieved LOS velocity displays a combination of upflows (redshifts; between $\log\tau_{500}\sim-4$ and $\log\tau_{500}\sim-3$) and downflows (blueshifts; beyond $\log\tau_{500}\sim-4$) in the stratified atmosphere. The uncertainties of the parameters, indicated by vertical bars at selected node positions, were evaluated using Eq.  (11.30) from \cite{2003isp..book.....D}, which takes into account both the discrepancy between the observed and the modeled profiles and the effective sensitivity of the spectral lines to the underlying parameters. Due to weak Stokes $Q$ and $U$ signals in \ion{Ca}{ii}~8542~\AA, the uncertainty in the inferred azimuth and transverse magnetic fields is higher in the chromospheric layers. 

\begin{figure}[!h]
\centering
\includegraphics[width=0.862\linewidth]{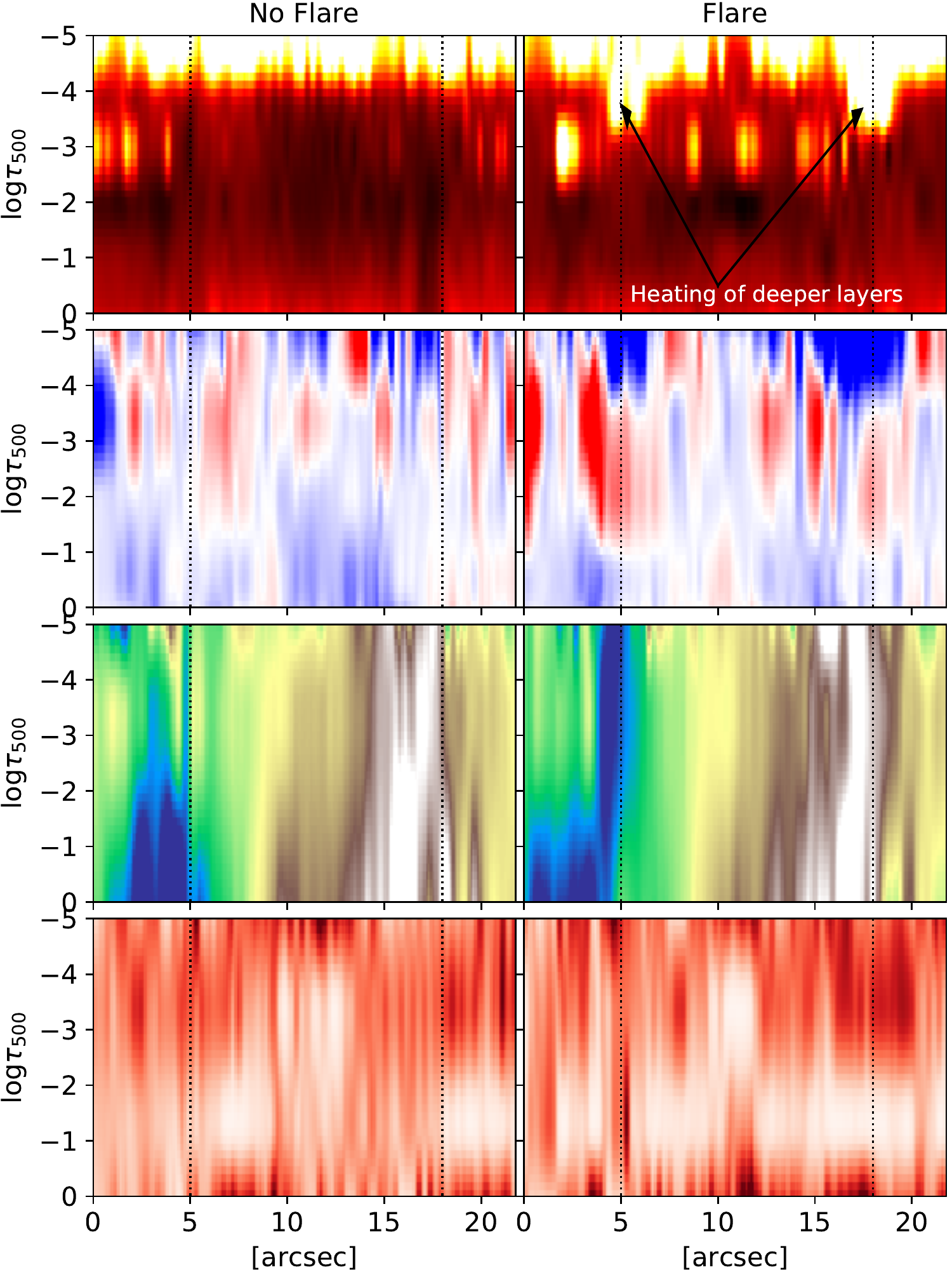}
\includegraphics[width=0.123\linewidth]{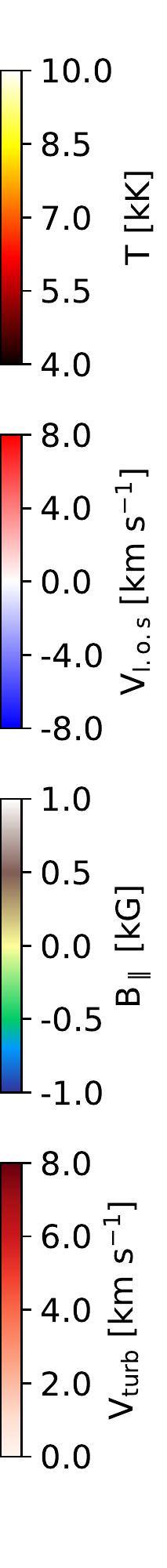}

\caption{\footnotesize Stratification of temperature, LOS velocity, LOS magnetic field, and microturbulent velocity obtained from the inversion of the pixels located on the dashed line shown in Fig.~\ref{fig_overview}. The left column shows the parameters retrieved around the no-flare time ($\sim$~09:30 UT), whereas the right column refers to the flare time ($\sim$~08:46 UT). The dotted vertical lines refer to the possible locations of the flare footpoints.}
\label{para_line}

\end{figure}

In addition to this, we also inverted the pixels located on the dashed line highlighted in Fig.~\ref{fig_overview}. The stratifications of inferred parameters, such as temperature, LOS velocity, magnetic field, and microturbulent velocity, are shown in Fig.~\ref{para_line}; they are evaluated at the flare peak time ($\sim$08:46~UT) and the no-flare time ($\sim$09:30~UT). The appearance of two opposite-polarity footpoints lying on the selected line is visible on the magnetic field map. The top panels of Fig.~\ref{para_line} clearly demonstrate the heating of the deeper layers, mainly at the flare footpoints locations. Moreover, the presence of chromospheric evaporation (upflows of hot plasma) and condensation (downflows of cooler plasma) near the footpoints are also remarkably visible. Furthermore, we find that the microturbulent velocity, which provides information about the unresolved and nonthermal component of the plasma velocity, is below 10~km~s$^{-1}$ in the flaring and non-flaring atmosphere. However, it is slightly higher in the flaring atmosphere. 
In the following section we further discuss the temporal evolution of the inferred parameters.


\begin{figure*}[!t]
\centering
\includegraphics[width=1\textwidth]{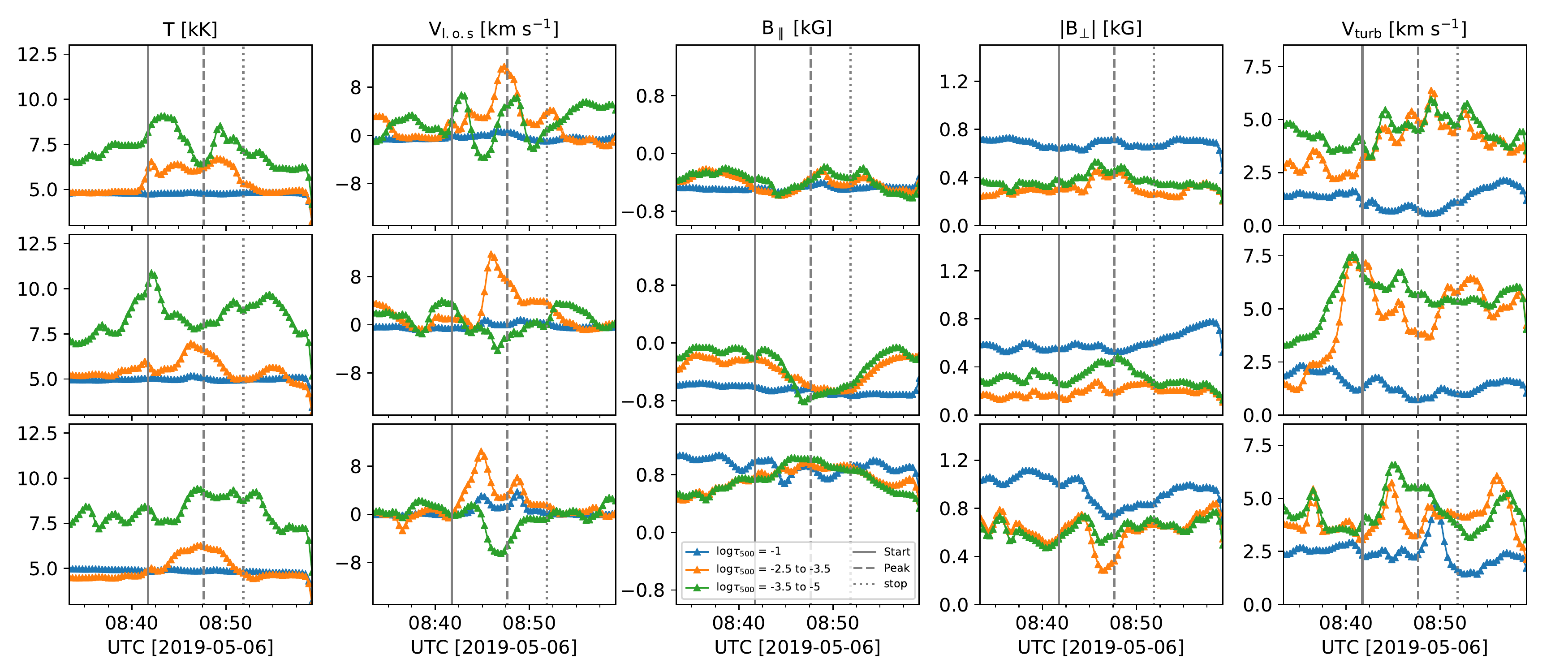}
\caption{\footnotesize Temporal evolution of the temperature, LOS velocity, longitudinal and transverse magnetic field, and microturbulent velocity obtained from the inversion with the STiC code at the P1 (top panels), P2 (middle panels), and P3 (bottom panels) locations highlighted in Fig. \ref{fig_overview}. The blue symbols refer to the mean values evaluated at $\log\tau_{500}\sim-1$. The orange and green symbols represent the mean values determined between $\log\tau_{500}\sim-2.5$ and $-3.5$ and between $\log\tau_{500}\sim-3.5$ and $-5$, respectively. The vertical solid, dashed, and dotted lines refer to the start, peak, and end times of the flare, respectively.}
\label{mean_para}

\end{figure*}

\subsection{Temporal evolution of the inferred parameters}
Figure \ref{mean_para} shows the temporal evolution of the parameters obtained from the inversion of Stokes profiles using the STiC code. At each timestep we obtained the mean value of the parameters for the selected pixels (5$\times$5) marked in Fig.~\ref{fig_overview}. For the sake of clarity and to enable comparison, we split the solar atmosphere into three regions based on the optical depth. The photospheric parameters were determined around the $\log\tau_{500}\sim-1$, whereas the mean values of the chromospheric parameters were evaluated between $\log\tau_{500}\sim-2.5$ and $-3.5$ for the lower chromosphere, and between $\log\tau_{500}\sim-3.5$ and $-5$ for the middle-to-upper chromosphere. 

At the beginning of the flare, the inferred temperature increases in the upper chromosphere and then decreases gradually after the flare end time. This is shown in the first column of Fig.~\ref{mean_para}. The mean value of the temperature increases from $\sim5$~kK up to $\sim11$~kK at the footpoint in the upper chromosphere, whereas in the lower chromosphere it changes from $\sim5$~kK to $\sim7$~kK. The temporal evolution analysis suggests that the flare heats the deeper layers (lower chromosphere) during its peak time, which is also evident in the top panels of Fig.~\ref{para_line}. Furthermore, the temperature in the lower chromosphere decreases to $\sim5$~kK just after the end time of the flare. However, in the upper chromosphere it decays more slowly when compared to the deeper layers. Our analysis suggests that the most intensively heated layer in the flaring atmosphere is the upper chromosphere, between $\sim7.5$~kK and $\sim11$~kK. In contrast to the chromosphere, we do not find a significant change in the photospheric temperature, as depicted by the blue curve in Fig. \ref{mean_para}.

The temporal evolution of the LOS velocity for the selected pixels is shown in the second column of Fig.~\ref{mean_para}. The layers between $\log\tau_{500}\sim-2.5$ and $-3.5$ exhibit downflows during the flaring time, with a maximum speed of $\sim12$~km~s$^{-1}$. In contrast, the middle-to-upper chromosphere ($\log\tau_{500}\sim-3.5$ to $-5$) shows upflows during the peak time near the flare footpoints, with a maximum value of $\sim-7$~km~s$^{-1}$. We notice that the upper chromosphere is dominated by upflows (evaporation) and the lower chromosphere by downflows (condensation), mainly at the flare footpoints. This is also illustrated in Fig.~\ref{para_line}, which shows the stratification of velocity at the flaring and non-flaring times. Our results are compatible with those reported by \cite{2002A&A...387..678F}, who also highlighted the presence of upflows and downflows in a flaring loop, but using the H$_{\delta}$, \ion{Ca}{ii}~K, and \ion{Si}{i}~3905~\AA\ spectral lines. 
In contrast, \cite{2017ApJ...846....9K} found weak downflows at an optical depth between $\log\tau_{500}\sim~-1$ and $\sim-5.5$, a discrepancy that could be attributed to observations of their C-class flare in its late phase and to the rather limited sensitivity of their observations in the \ion{Ca}{ii}~8542~\AA\ line to the photospheric velocities. In the photosphere, we do not notice a significant change in the LOS velocity, which is in agreement with various studies \citep{2002A&A...387..678F, 2017ApJ...846....9K,Kuridze2018}.

The temporal evolution of the LOS magnetic field obtained from the inversions is shown in the third column of Fig.~\ref{mean_para}. In our analysis we do not find permanent or step-wise change in B$_{\rm LOS}$ from pixel to pixel.
However, the inferred B$_{\rm LOS}$ in the lower and upper chromospheres displays changes during the flaring time at the flare footpoints located at opposite polarities. As depicted in Fig.~\ref{mean_para} for the pixels located on the negative polarity footpoint, the change begins $\sim4$~minutes after the flare start time, increases until the flare peak time from $\sim$150~G to $\sim$800~G, and then decreases again to near the pre-flare B$_{\rm LOS}$ value ($\sim~150$~G) for both the upper and lower chromospheres. We also notice similar B$_{\rm LOS}$ changes in the positive polarity footpoint pixel, which exhibits a change from $\sim$500~G to $\sim$1000~G. In the chromosphere, we do not notice any significant change in the horizontal component of the magnetic field (shown in the fourth column of Fig.~\ref{mean_para}), which could be due to noisy Stokes~$Q$ and $U$ signals. We note that the inferred horizontal magnetic field B$_{\perp}$ and azimuth angles show more uncertainties in the chromosphere due to weak signals in the Stokes~$Q$~and~$U$ signals. In contrast to the chromospheric LOS magnetic field, we do not notice any significant changes in the photospheric LOS magnetic field. Noticeably, in the photosphere, the pixels located near the footpoint of positive polarity exhibit a small change of $\sim$200~G in the horizontal magnetic field strength, whereas the pixels located near a negative polarity footpoint exhibit a small linear increase after the peak time.

The obtained microturbulent velocity is also shown in the fifth column of Fig.~\ref{mean_para}. It illustrates that the V$_{\rm turb}$ value is variable throughout the flaring atmosphere in the chromosphere, whereas in the photosphere it remains below $\sim2.5$~km~s$^{-1}$. We notice a significant rise in the V$_{\rm turb}$ value near the flare start time. During the flaring time the value is between $\sim2.5$~km~s$^{-1}$ and $\sim7.5$~km~s$^{-1}$. 

\begin{figure*}[!ht]
\centering
\includegraphics[width=0.325\textwidth]{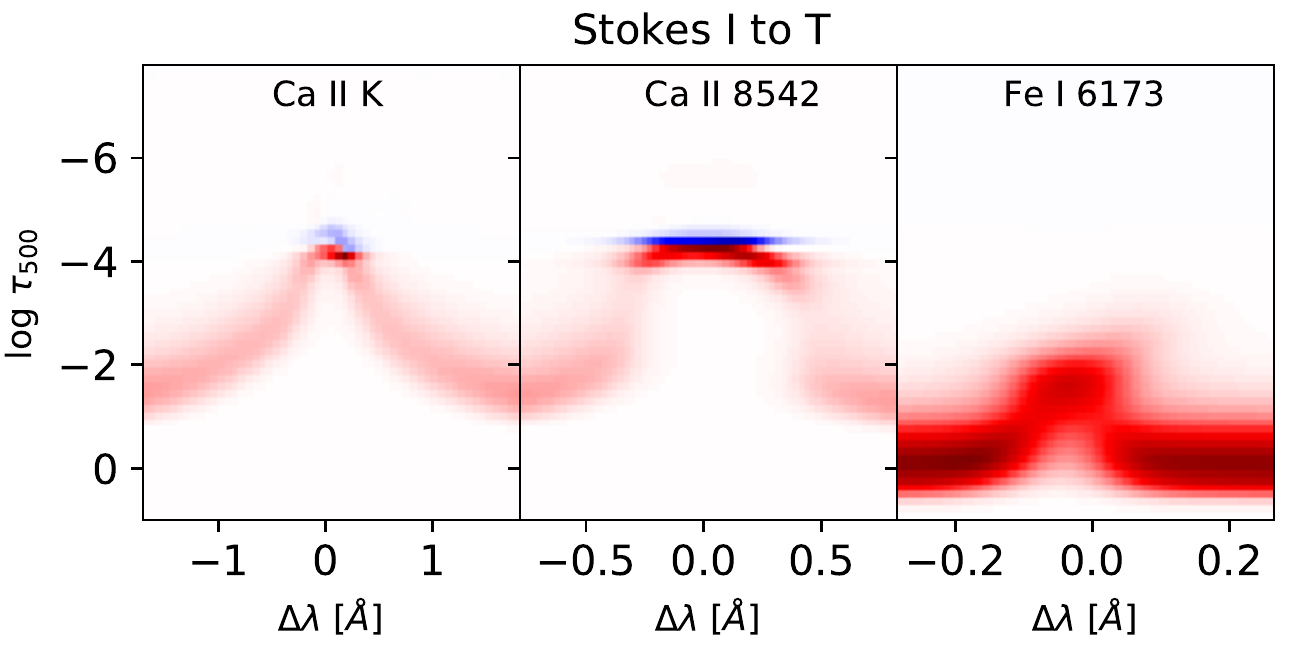}
\includegraphics[width=0.325\textwidth]{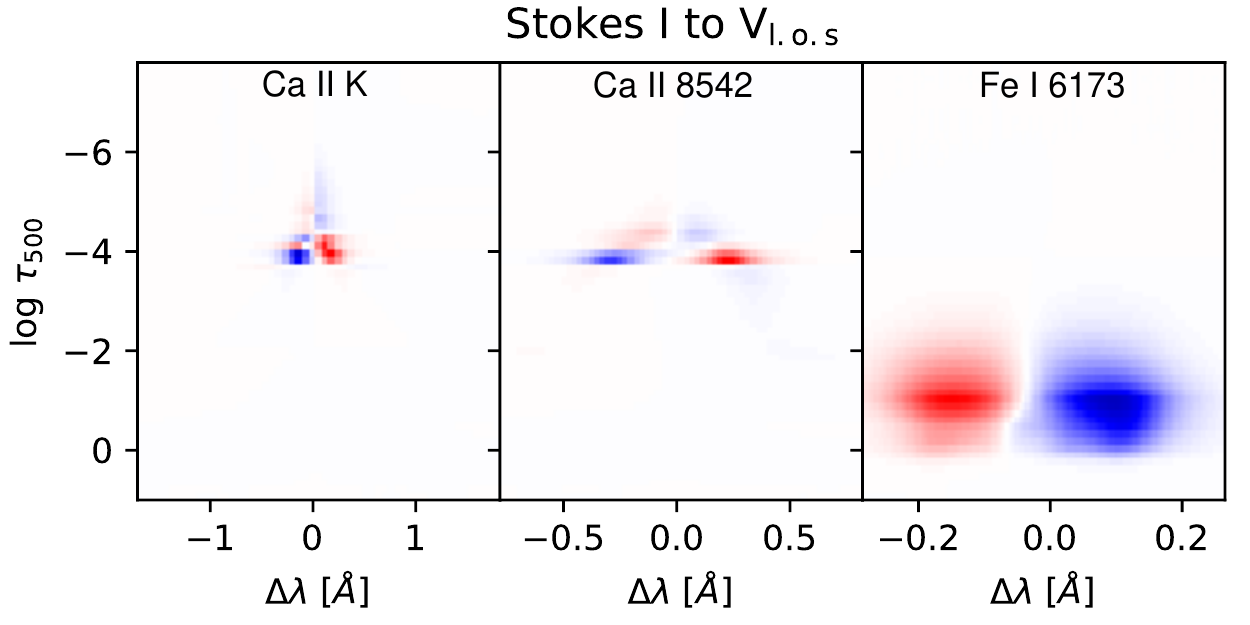}
\includegraphics[width=0.325\textwidth]{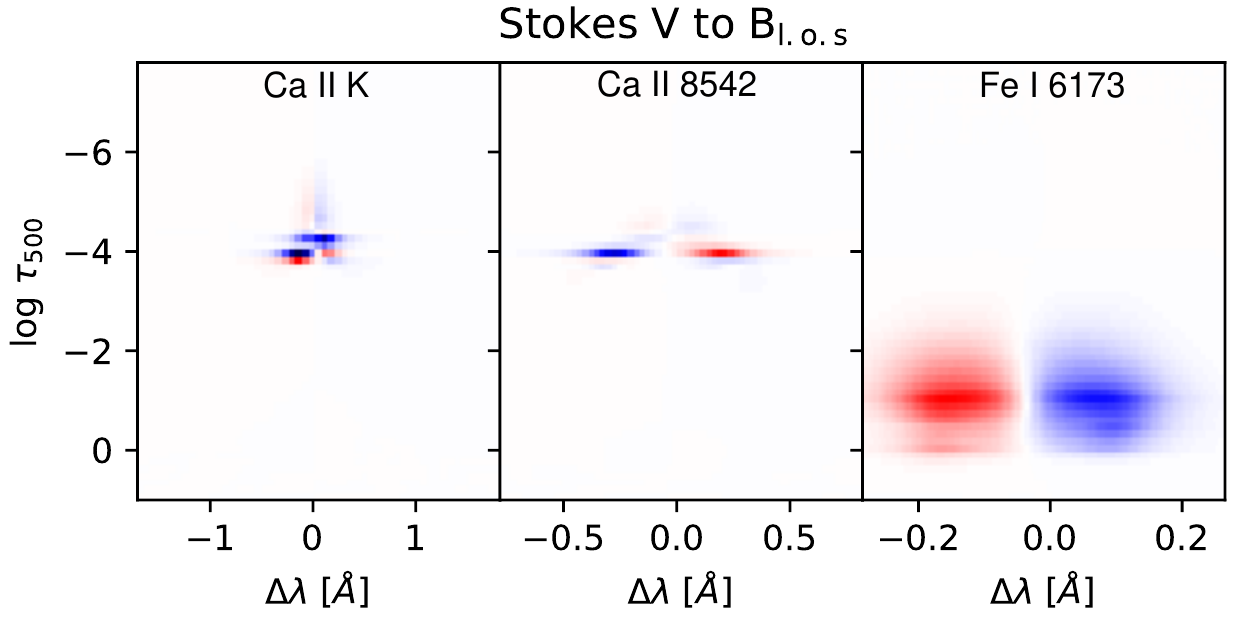}
\includegraphics[width=0.325\textwidth]{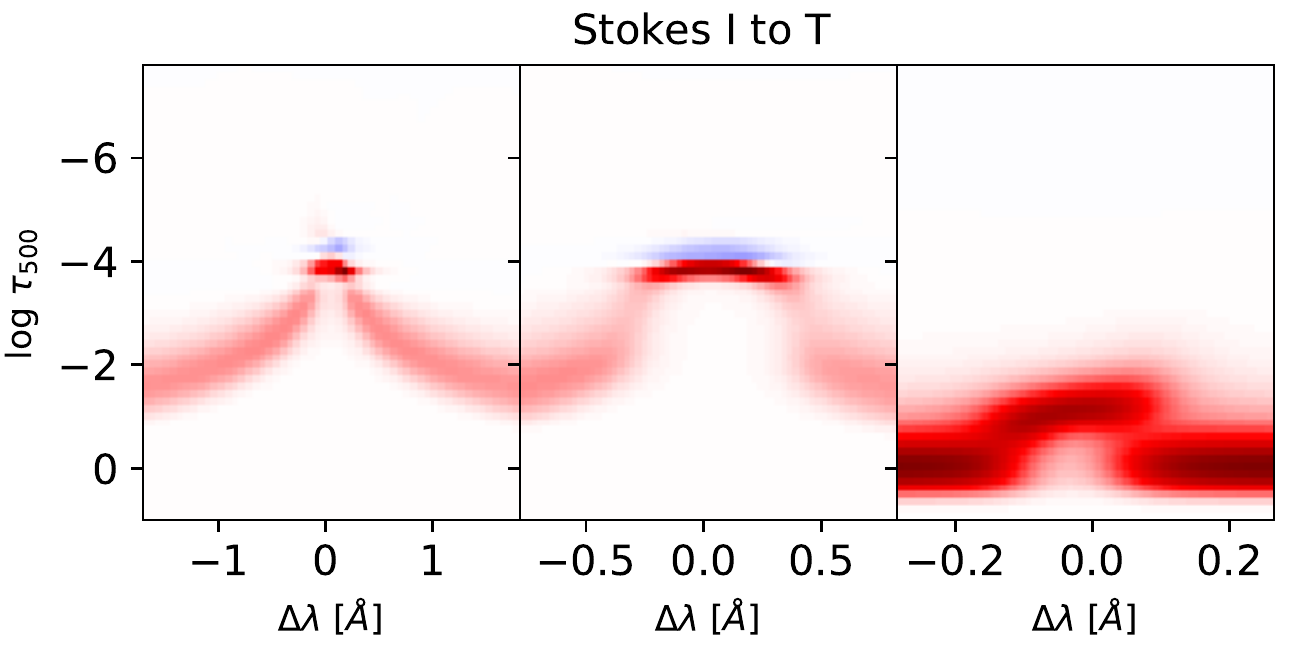}
\includegraphics[width=0.325\textwidth]{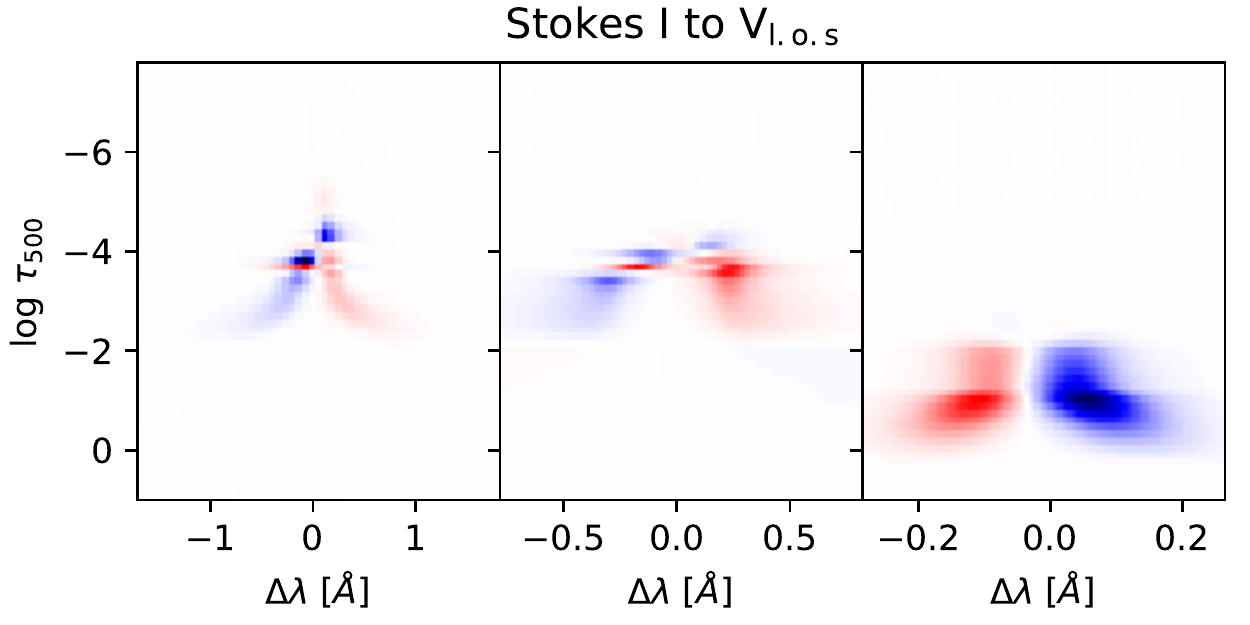}
\includegraphics[width=0.325\textwidth]{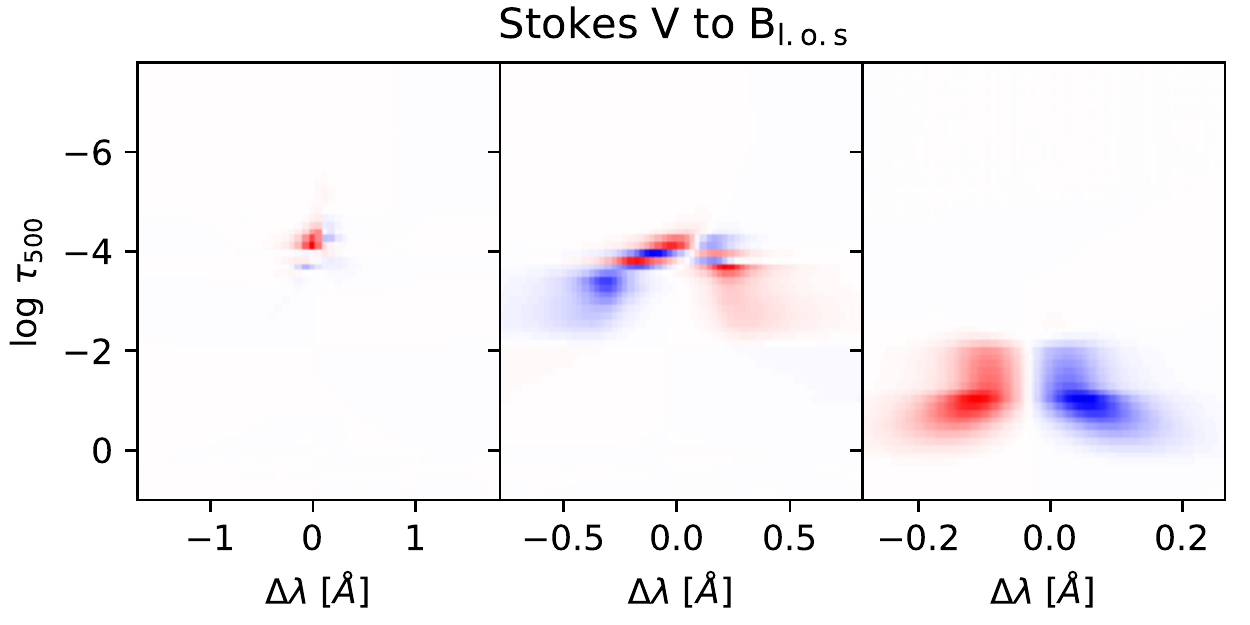}
\includegraphics[width=0.495\textwidth]{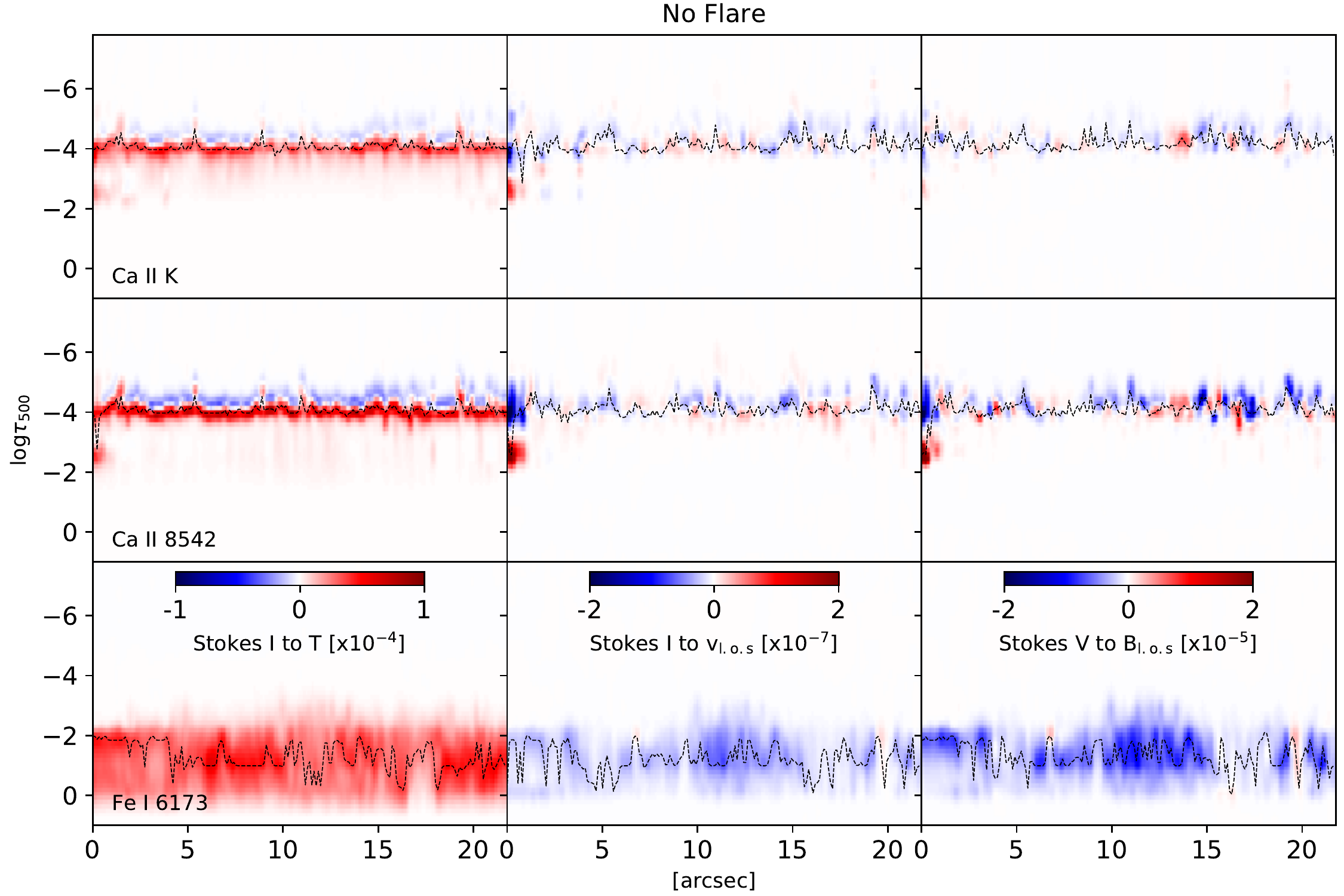}
\includegraphics[width=0.495\textwidth]{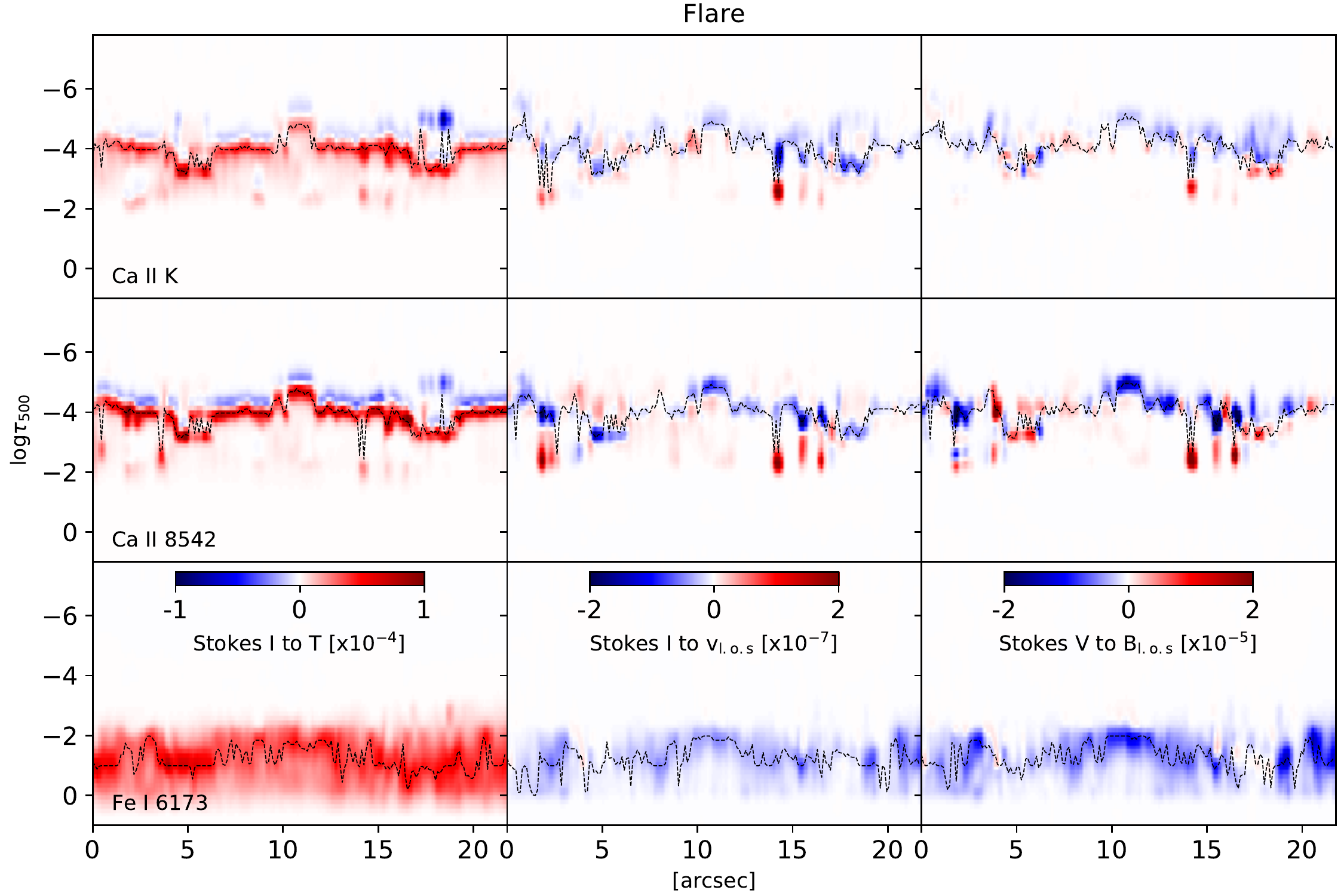}
\caption{\footnotesize Response functions of the Stokes profiles at the flare and no-flare time. Top and middle rows: Response functions of the Stokes profiles to temperature, velocity, and LOS magnetic field as a function of wavelength, obtained during the no-flare time (top panel) and the flare peak time (middle panel). Bottom panels: Mean RFs around the line core for the pixels located on the dashed line at the flare time (right) and the no-flare time (left). The dotted black lines refer to the maximum RF values.}
\label{rf_line}
\end{figure*}

\begin{figure}[!h]
\centering
\includegraphics[width=1\linewidth]{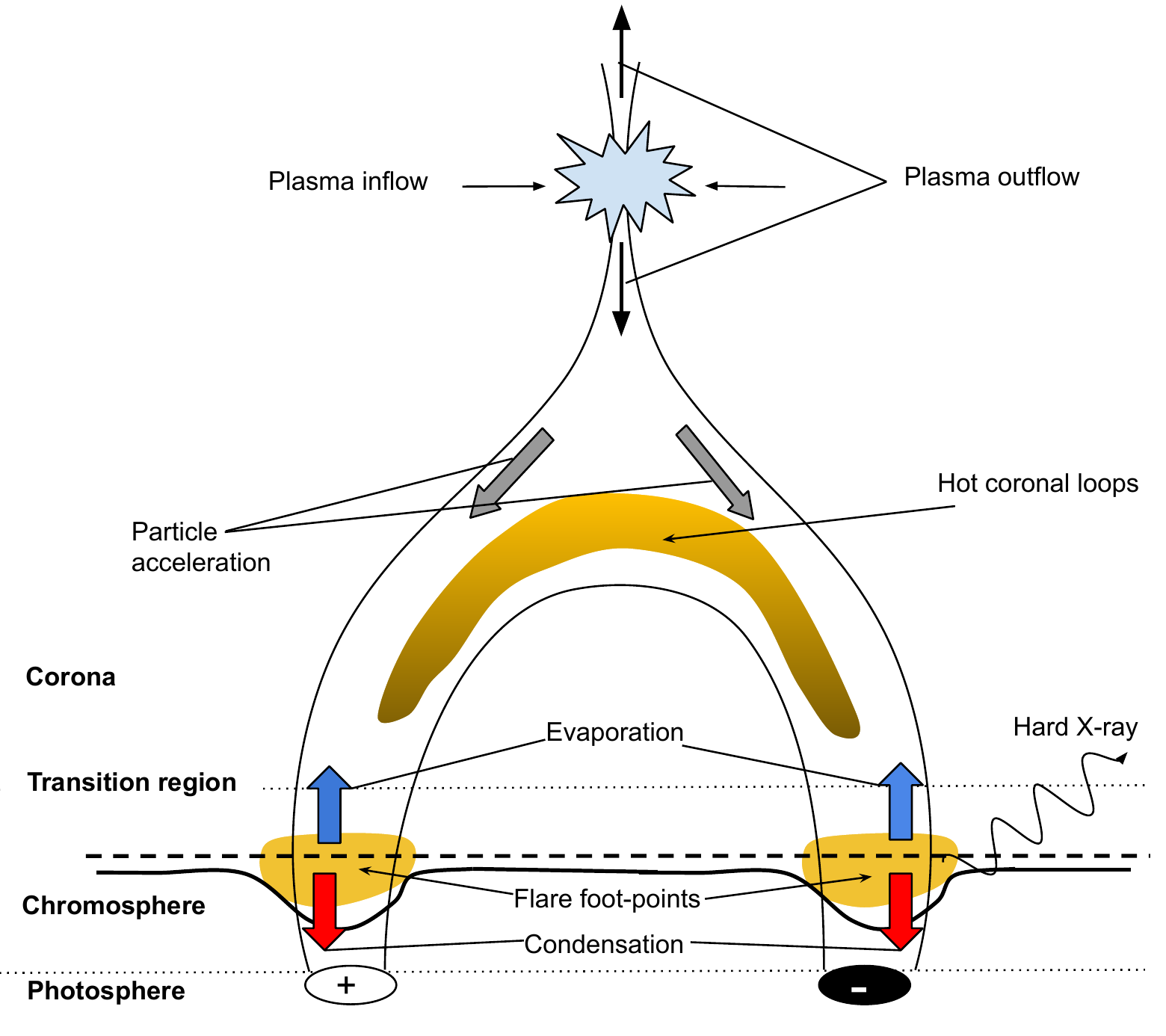}
\caption{\footnotesize Sketch of the standard flare model. A plausible scenario in the chromosphere explaining our observations is depicted. The observed chromospheric condensation and evaporation near the flare footpoints are indicated by red and blue arrows, respectively. The dashed black line and the solid black line refer to the average RFs (around the line core) of the chromospheric lines (\ion{Ca}{ii}~K \& \ion{Ca}{ii}~8542~\AA) to the temperature during the no-flare and flare peak times, respectively.}
\label{cartoon}

\end{figure}

\subsection{Sensitivity of the spectral lines}
\label{sec_response}
In this section we discuss how different lines (\ion{Ca}{ii}~K, \ion{Ca}{ii}~8542~\AA, and \ion{Fe}{i}~6173~\AA) are sensitive to the flaring atmosphere. Response functions (RFs) provide valuable information about how the Stokes profiles respond to perturbations in the physical parameters of the model atmosphere, such as temperature, magnetic field, LOS velocity, etc. \citep{1975SoPh...43..289B,2003isp..book.....D}. Additionally, they are also used to compute the uncertainty in the inferred physical parameters.

We evaluated the RFs for the pixels located on the dashed line shown in Fig.~\ref{fig_overview}, covering the flare footpoints and the PIL region. The stratification of inferred parameters along this line are also shown in Fig.~\ref{para_line}. In order to compare the flaring and non-flaring atmospheres, the physical parameters were evaluated at the flare peak time ($\sim~$08:46~UT) and at a later (no-flare) time ($\sim~$09:30~UT).

As an example, the RFs of the Stokes profiles to the physical parameters as a function of wavelength for a flaring pixel and a non-flaring pixel are shown in the top and middle panels of Fig.~\ref{rf_line}, while the bottom panels display the mean RFs around the line core along the dashed line highlighted in Fig.~\ref{fig_overview}. We took the LOS velocity variation into account  while evaluating the mean RFs around the line core.

As depicted in Fig.~\ref{rf_line}, the RFs of Stokes~$I$ to the temperature, for the \ion{Ca}{ii}~K and 8542~\AA\ lines, exhibit sensitivity between $\log\tau_{500}\sim-1$ and $-5$ for both the non-flaring and flaring pixels. However, the sensitivity around the flare peak time moves toward the deeper layers. We note that the transition region in the reconstructed models moves accordingly to deeper layers as well, which is compatible with results from numerical radiative hydrodynamic simulations \citep[e.g.,][]{2016ApJ...827..101K}. Furthermore, the RFs of Stokes~$I$ to LOS velocity show maximum sensitivity around $\log\tau_{500}\sim-4$ for a non-flaring pixel, whereas for the flaring pixels it becomes sensitive to deeper layers, down to $\log\tau_{500}\sim-2.5$. The RFs of Stokes~$V$ to the LOS magnetic field for the \ion{Ca}{ii}~8542~\AA\ line show more sensitivity in the deeper layers ($\log\tau_{500}\sim-2.5$ to $-4$) in the flaring pixels compared to the non-flaring pixels, which are mainly only sensitive around $\log\tau_{500}\sim-4$.  

It is clear from the obtained mean RFs of Stokes profiles to the temperature, shown in the bottom panels of Fig.~\ref{rf_line}, that in the flaring atmosphere the \ion{Ca}{ii}~K and \ion{Ca}{ii}~8542~\AA\ lines show more sensitivity in the deeper layers of the atmosphere, mainly at the footpoints locations ($\sim$6\arcsec and $\sim$17\arcsec). This suggests that even a C-class flare can heat the middle-to-lower chromosphere similarly to the intense M- and X-class flares. It also provides confidence in the reliability of the inferred physical parameters in the deeper layers of the atmosphere. Similar to the temperature, we also notice changes in LOS velocity and magnetic field, mainly at the locations of the flare footpoints. Furthermore, the response of the Stokes~$V$ profiles to the LOS magnetic field also moves toward deeper layers during the flaring time compared to the non-flaring time. 

In contrast to the \ion{Ca}{ii} lines, we do not notice significant change in the RFs for the \ion{Fe}{i}~6173~\AA\ line between the flaring and the non-flaring pixels, which implies that the photosphere is not affected by our C-class flare.

\section{Discussion}
\label{sec_conclusion}

In this study, we have analyzed a C2-class solar flare (SOL2019-05-06T08:47) that occurred on May 6, 2019, in NOAA AR 12740. The flare was observed simultaneously in the \ion{Ca}{ii}~K, \ion{Ca}{ii}~8542~\AA,\ and \ion{Fe}{i}~6173~\AA\ lines with the CRISP and CHROMIS instruments at the SST.
We performed full Stokes imaging spectropolarimetric observations in the \ion{Ca}{ii}~8542~\AA\ and \ion{Fe}{i}~6173~\AA\ lines, whereas we observed only Stokes~$I$ in the \ion{Ca}{ii}~K spectral line. Our observations cover the rise, peak, and decay phases of the flare.
All observed lines were analyzed simultaneously in order to understand the stratification of the physical parameters, mainly at the flare footpoints, in the flaring and non-flaring atmospheres. In the following, we summarize our results.

The temporal evolution clearly shows the presence of intense brightening near the flare footpoints. 
At these locations the observed Stokes~$I$ profiles of \ion{Ca}{ii}~K and \ion{Ca}{ii}~8542~\AA\ are broad, asymmetric, and show strong emission peaks during the flaring  time. 
Our analysis suggests that the asymmetries in the \ion{Ca}{ii} lines are due to the presence of strong gradients in the LOS velocity. It is known that flares mainly heat the chromosphere at the flare footpoints, which can produce a reversal in the gradient of the source function, leading to an increase in the line core emission relative to the far wings. As a result, strong emission profiles in the chromospheric lines are expected, and are indeed present in our observations.

The temporal analysis of the LOS magnetic flux near the PIL and the flare footpoints reveals that both the longitudinal and the transverse magnetic fluxes exhibit changes during the flare in the photosphere as well as in the chromosphere. These chromospheric changes are correlated with the GOES X-ray curve, which suggests that reconfiguration of the magnetic field takes place in the upper atmosphere. In contrast to previous studies \citep{2010ApJ...724.1218P,2017ApJ...834...26K,2018ApJ...852...25C}, we do not find any step-wise-like change in the pixels located near the flare footpoints and the PIL. This could be attributed to a less intense C-class flare. Noticeably, we find more abrupt changes in the LOS magnetic field in the chromosphere as compared to the photosphere around the flare peak time.

In order to analyze the stratification of the flaring atmospheres, we employed the non-LTE multiline inversion code STiC. We investigated the temperature, the LOS magnetic field, the LOS velocity, and the microturbulent velocity for a few selected pixels, mainly located at the flare footpoints and at the PIL. 
The temporal analysis of the flare footpoints shows that the upper chromosphere ($\log\tau_{500}\sim-3.5$ and $-5$) is heated between $\sim7.5$~kK and $\sim11$~kK, whereas the lower chromosphere ($\log\tau_{500}\sim-2.5$ to $-3.5$) is heated up to 7~kK.

In the photosphere (i.e., below $\log\tau_{500}\sim-2$) we do not notice any significant change during the flare and non-flare times. The obtained temperature stratification during a C2-class flare is consistent with \cite{Kuridze2018}; however, they reported a higher temperature value up to $\sim~$13~kK, which could be due to the stronger intensity of the M1.9-class flare. Moreover, and in contrast with the present study, \cite{2017ApJ...846....9K} reported higher temperature values ranging between $\sim$6.5~kK and $\sim$20~kK in the middle and upper chromospheres ($\log\tau_{500}\sim-3.5$ to $-5.5$) for a C8.4-class flare. Normally, at such high temperatures, the \ion{Ca}{ii} is completely ionized, leaving higher-order ionized atoms. This overestimated temperature value could be attributed to the lack of sensitivity ensuing from the inversion of a single spectral line (\ion{Ca}{ii}~8542~\AA) or to the interpolation approach between the selected nodes for temperature, which could lead to an extrapolation of the temperature gradient from the lower chromosphere into the higher layers where the line is no longer sensitive \citep[e.g.,][]{daSilva2018}.

The obtained LOS velocity has revealed a combination of upflows and downflows during the flaring time at the flare footpoint locations, which are the signatures of the chromospheric condensation and evaporation. We find that the lower chromosphere ($\log\tau_{500}\sim-2.5$ to $-3.5$) exhibits downflows, with a maximum value of $\sim12$~km~s$^{-1}$, whereas the upper-to-middle chromosphere ($\log\tau_{500}\sim-3.5$ to $-5.5$) hosts upflows of $7$~km~s$^{-1}$. 
In addition to flare observations \citep[e.g.,][]{2002A&A...387..678F, 2008A&A...490..315B,2019A&A...621A..35L}, radiation hydrodynamics simulations \citep[e.g.,][]{2016ApJ...827..101K,2020ApJ...895....6G} and 3D radiative magnetohydrodynamic simulations \citep[e.g.,][]{2019NatAs...3..160C} of solar flares also report similar scenarios of upflows and downflows.


The temporal analysis of the LOS magnetic field inferred from the STiC inversion code at the footpoints exhibits changes during the flaring time. Both opposite-polarity footpoints display an increase in the magnetic field strength at the flare peak time. 
The retrieved maximum change at the footpoints is around 650~G. 

To investigate the sensitivity of the observed lines in the flaring atmosphere, we analyzed the RFs computed for the model atmosphere obtained from the STiC inversion code. To facilitate the comparison, the numerical RFs were computed at the flare peak time ($\sim$08:46 UT) and after the flare ended ($\sim$09:30 UT) across a line passing through the flare footpoints and the PIL. The analysis of average RFs to temperature near the line core of the \ion{Ca}{ii} and \ion{Fe}{i}~6173~\AA\ lines in the flaring atmosphere shows that the \ion{Ca}{ii} lines are more sensitive in the deeper layers ($\log\tau_{500}\sim-$2.5 to $-$4) compared to the non-flaring atmosphere, where they are mainly sensitive around $\log\tau_{500}\sim-$4. Similar to the temperature, the deeper layers are more sensitive to the LOS velocity and LOS magnetic field, which is in agreement with \cite{Kuridze2018}. In the photosphere, below $\log\tau_{500}\sim-$2.5, we do not notice a significant difference between the flaring and non-flaring atmospheres. 

Our RF analysis suggests that a fraction of the apparent increase in the LOS magnetic field, mainly at the footpoints, may be due to the increase in the sensitivity of the \ion{Ca}{ii}~8542~\AA\ line in the deeper layers, where the field is relatively strong compared to the upper layers. The rest may be due to the reconfiguration of magnetic field lines during the flare.

Various models have been proposed to understand a solar flare (see \citealt{2011LRSP....8....6S}, and references therein). In the standard flare model, commonly known as the ``CSHKP'' model \citep[originally introduced by][]{1964NASSP..50..451C,1966Natur.211..695S,1974SoPh...34..323H,1976SoPh...50...85K}, the flare is triggered by the magnetic reconnection or reconfigurations of the magnetic field lines in the corona. During magnetic reconnection processes, the magnetic energy is converted into thermal and kinetic energy, resulting in plasma heating, particle acceleration, and a release of energy in the entire electromagnetic spectrum. A beam of energetic particles is accelerated to the deeper layers of the solar atmosphere, where they deposit most of their energy and momentum. The interaction with the dense chromospheric plasma gives rise to heating and the production of hard X-ray and soft X-ray emissions. At the flare footpoints the hot plasma rises (evaporation) and the cool plasma sinks (condensation) toward the lower atmosphere along the flare loops.

Unfortunately, due to the unavailability of X-ray observations, we cannot provide evidence of hard X-rays at the flare footpoints. However, recent X-ray observations of flares have demonstrated that, normally, hard X-rays are emitted near the flare footpoints \citep{2002SoPh..210..307F,2011ApJ...735...42B,2015ApJ...813..113B}. In line with the standard flare model, the presence of both chromospheric evaporation and condensation at the flare footpoints is clearly noticed in our observations. In order to present a possible scenario of the flare, we sketch the magnetic field configuration and the observed features together with the standard flare model in Fig. \ref{cartoon}. This sketch illustrates the simplest picture of the flare by considering the standard flare model in 2D and the results of this study. It depicts the appearance of chromospheric evaporation and condensation at the flare footpoints and the mean RFs of the chromospheric lines to the temperature for the flaring and non-flaring atmospheres.


\section{Conclusions}

In the present paper we have reconstructed the stratification of flaring atmospheres as a function of time using a multiline full-Stokes inversion approach. Our main results can be summarized as follows.
\begin{itemize}
    \item In the footpoints we reconstruct the simultaneous presence of chromospheric condensation and evaporation. At that location the temperature rises up to approximately 11~kK.
    \item The sensitivity of the \ion{Ca}{II} lines analyzed in this study shifts to a larger optical depth in the footpoints and their surroundings. However, the \ion{Fe}{i}~6173 \AA\ line shows insignificant changes during the flare.
    \item Our time evolution analysis yields changes in the magnetic field stratification, mostly above the photosphere, but those changes are not step-wise, contrary to what was reported in previous studies involving more intense flaring conditions.
\end{itemize}

It has been speculated that the magnetic field lines reconfigure in the upper layers during a flare. The direct signatures of magnetic field changes derived from flare observations are very limited. Thus, to further investigate this hypothesis, we need simultaneous photospheric and chromospheric polarimetry with sufficiently strong linear polarization signals, which can be achieved with state-of-the-art instruments and the advent of new generation solar telescopes, such as the Daniel K. Inouye Solar Telescope (DKIST; \citealt{2015csss...18..933T}) and the European Solar Telescope (EST; \citealt{2016SPIE.9908E..09M}).
Our future work will focus on the analysis of multiline polarimetric observations to understand the changes in the  magnetic field vector during a flare.

\begin{acknowledgements}
The Swedish 1-m Solar Telescope is operated on the island of La Palma by the Institute for Solar Physics of Stockholm University in the Spanish Observatorio del Roque de los Muchachos of the Instituto de Astrof\'isica de Canarias. The Institute for Solar Physics is supported by a grant for research infrastructures of national importance from the Swedish Research Council (registration number 2017-00625).
RY and FC are supported through the CHROMATIC project (2016.0019) funded by the Knut och Alice Wallenberg foundation.
JdlCR is supported by grants from the Swedish Research Council (2015-03994), the Swedish National Space Board (128/15) and the Swedish Civil Contingencies Agency (MSB). This project has received funding from
the European Research Council (ERC) under the European Union's Horizon 2020 research and innovation program (SUNMAG, grant agreement 759548).
The inversions were performed on resources provided by the Swedish National Infrastructure for Computing (SNIC) at the High Performance Computing Center at Link\"oping University.
Data and images are courtesy of NASA/SDO and the HMI and AIA science teams. This research has made use of NASA’s Astrophysics Data System.
We acknowledge the community effort devoted to the development of the following open-source packages that were used in this work: numpy (\url{numpy.org}), matplotlib (\url{matplotlib.org}) and sunpy (\url{sunpy.org}).
\end{acknowledgements}

\bibliographystyle{aa}
\bibliography{new-ref}  

\end{document}